\documentclass[11pt]{article}
\usepackage{multirow}
\usepackage{fullpage}
\usepackage[utf8]{inputenc}
\usepackage{amsthm}
\usepackage{amsmath}
\usepackage{amssymb}
\usepackage{amsfonts,enumerate}
\usepackage[dvipsnames]{xcolor}
\usepackage{hyperref}
\usepackage{graphicx}
\usepackage{tikz}
\usepackage{caption}
\usepackage{subcaption}
\usepackage{authblk}

\usetikzlibrary{shapes.geometric, arrows}
\usepackage[
backend=biber,
style=numeric,
]{biblatex}
\renewbibmacro{in:}{}

\DeclareMathOperator{\atts}{Y_\text{ATT}^\text{S}}
\DeclareMathOperator{\E}{E}

\newcommand{\ind}{\perp\!\!\!\!\perp} 
\DeclareMathOperator{\N}{\mathcal{N}}
\DeclareMathOperator{\sign}{sign}
\DeclareMathOperator{\R}{\mathbb{R}}
\DeclareMathOperator{\prob}{\mathbb{P}}
\DeclareMathOperator{\hnull}{\mathcal{H}_0}

\DeclareMathOperator{\law}{\mathcal{L}}

\newtheorem{theorem}{Theorem}
\newtheorem{proposition}{Proposition}

\newtheorem{lemma}{Lemma}

\newif\ifshowcomments
\showcommentstrue

\title{Mapping beyond diseases: Controlled variable selection for secondary phenotypes using tilted knockoffs}

\author[1]{Qian Zhao}
\author[2]{Susan Service}
\author[2, 3]{Carrie E. Bearden}
\author[4]{Carlos Lopez-Jaramillo}
\author[2]{ Nelson Freimer}
\author[5]{Chiara Sabatti}

\affil[1]{Department of Mathematics and Statistics, University of Massachusetts Amherst }
\affil[2]{Center for Neurobehavioral Genetics, Semel Institute for Neuroscience and Human Behavior, University of California Los Angeles}
\affil[3]{Department of Psychology, University of California Los Angeles}
\affil[4]{Department of Psychiatry, Universidad de Antioquia}
\affil[5]{Departments of Biomedical Data Science and Statistics, Stanford University}

\begin{document}
\maketitle

\section{Introduction}\label{section:introduction}

This work is motivated by a study \cite{service2020} that aims to identify genetic underpinnings of  severe mental illnesses (SMI)---schizophrenia, bipolar disorder and major depressive disorder---exploring not only the major diagnoses, but also their sometimes overlapping symptoms and correlated outcomes. An ongoing study in the Paisa region in Colombia led to the genotyping of a case-control sample for SMI, as well as the assessment of detailed symptomatology and neurocognitive functions \cite{service2020}. Researchers recruited over 8,800 individuals in the study, among whom 13\% are diagnosed with schizophrenia, 33\% are bipolar disorder and 31\% major depressive disorder. Endophenotypes collected in the study include forty symptoms in the domains of psychosis, depression and mania, and nine tests from the Computerized Neurocognitive Battery (CNB) measuring executive function, episodic memory, complex cognition, social cognition, and motor praxis.
% who are patients who have been hospitalized because of SMI, and controls, who are healthy individuals living in the same community. 
Given the challenges presented by diagnosis of SMI, and how it might vary across cultures, one research direction  is the identification of  genetic variants associated with these endophenotypes, measurable traits that are linked to an illness and are thought to be more closely related to genes than clinical diagnoses \cite{Bearden_Freimer_2006}. Recent studies have shown that studying genetic underpinnings of endophenotypes that mediate disease outcomes may improve prediction accuracy of polygenic risk scores \cite{Kharitonova2024}.  Another area of interest is whether there are direct pathways between genetic variation and neurocognitive functions (e.g., attention or working memory), beyond the impact on these of SMI. 
%The second question is of interest because different SMI often overlap in symptoms as well as genetic variants, suggesting that subsets of genetic variants may collectively impact some endophenotypes, which in turn affects the diagnosis. Therefore, studying these secondary phenotypes might provide biological insights which might lead to more precise diagnosis and treatment. 
Two statistical challenges emerge. First, because the distribution of the case-control sample differs from the population, we might observe spurious associations between genetic variants and the secondary phenotype.  Secondly, because we test hundreds and thousands of genetic variants at the same time, we want put in place a procedure that corrects for multiplicity, by controlling the false discovery rate (FDR). 

While in the interest of concreteness, we have focused on this particular study \cite{service2020}, the problem that we seek to address is quite widespread. Genetic studies very often collect data on the secondary phenotypes that can  elucidate biological pathways underlying diseases; and in practice, the samples  researchers work with are biased. For instance, in a case-control study, cases are intentionally over-sampled in order to study risk factors of rare diseases; bio-bank data may exhibit sampling bias caused by individuals self-selecting into the study \cite{griffith2020CovidCollider, pirastu2021sex,participationBiasUKBB_schoeler2023}. In both situations, a researcher might observe a spurious association between a genetic variant and a trait if both affect the selection probability. This phenomenon, often referred to as ``collider bias'' (or ``index event bias'', ``reversal paradox''), has recently been brought to attention in empirical studies, and researchers have reported spurious associations as well as biased estimates as a result  \cite{griffith2020CovidCollider,pirastu2021sex,akimova2021ColliderPRS, Day2016}. To address the general problem of collider bias, statisticians have proposed several methods to obtain unbiased estimates \cite{Yaghootkar2017,groenwold2019,linSPML2009,ruchardson2007ipw,xing2016robust} or test a hypothesis \cite{tenzer2022}. Previous research, however, focused on testing association between a pair of variables, while in genome-wide association studies (GWAS) it is necessary to explore the importance of a large number of variables, and to put in place strategies to control false discoveries. This article seeks to address this challenge leveraging the knockoff framework \cite{candes2018modelX}.

\subsection{Problem setting}\label{section:problem_setting}

\begin{figure}[ht]

\begin{tikzpicture}
% the populatin diagram
% add a bounding box outside of the population

% specific example 

\node[](population) at (5.5,2.3) {(a) };
\node[](population) at (13.5,2.3) {(b) };

\node[anchor=west](population) at (4.8,0) {{\bf Population} }; %, distribution  $ \prob({X},Y)$, hypotheses ${\cal H}_0$};
\node[anchor = west](population2) at (7.2,0) {$ \prob({X},Y)$}; %, distribution  $ \prob({X},Y)$, hypotheses ${\cal H}_0$};

\node[](SNP1) at (0,0) {${X}_1$};
\node[](SNP2) at (1,0) {${X}_2$};
\node[](dots) at (2.2,0) {$\ldots$};
\node[](SNPp) at (3.2,0) {${X}_p$};

\node[](outcome2) at (4.5,0) {$Y$};
%\node[](Disease) at (-1,-1.5) {Diagnosis, driving selection};
%\node[](D) at (5,-1.5) {$D$};
\node[anchor = west](sample) at (4.8,-2.5) {{\bf Sample}}; %, distribution ${\cal L}( {X},Y)$};
\node[anchor = west](sample2) at (7.2,-2.5) {${\cal L}( {X},Y)$}; %, distribution ${\cal L}( {X},Y)$};

\node[](S2) at (2.3,-2.5) {$S$};
%\node[](fig2) at (8, -3) {(b)};

\draw [dashed, line width=0.8 pt] (0.2, 0.2) .. controls (2.2, 0.8) .. (4.3,0.2);
 \node[](question1) at (2.2, 1) {? };
\node[](question1) at (4.2, 1.5) {\quad Object of inference \quad $\{{\cal H}_{0j}:X_j\ind Y \,|\, X_{-j}\}_{j=1}^p$};

%\draw [->] (SNP1) -- (D);
%\draw [->] (SNP2) -- (D);
%\draw [->] (SNPp) -- (D);
%\draw [->] (outcome2) -- (D);

\draw [->] (SNP1) -- (S2);
\draw [->] (SNP2) -- (S2);
\draw [->] (SNPp) -- (S2);
\draw [->] (outcome2) -- (S2);
% \draw [->] (D) -- (S2);

% add illustration of the sampling mechanism 
\draw[thick, dashed] (10, 1) rectangle (16,-3);

% add text about the population 
\node[](population_box) at (14.6, .5) {Population};
% add text about the case control sample
\node[](sample_box) at (11, -2.5) {Sample};

% add illustration of the case control sample
\filldraw[fill=black!50!white, draw=black] (15,1.6) circle (0.1cm);
\draw (12.8,1.6) circle (0.1cm);

\filldraw[fill=black!50!white, draw=black] (15, 1.6 -.1) --  (15 -0.15, 1.6 - 0.4)-- (15 +0.15, 1.6- 0.4) -- cycle; 
\draw (12.8, 1.56- 0.1) --  (12.8 -0.15, 1.6 - 0.4)-- (12.8+0.15, 1.6 - 0.4) -- cycle; 
\node[anchor=east]() at (14.7, 1.5) {Cases};
\node[anchor=east]() at (12.5, 1.5) {Controls};

% add another box outside the two distribution, one linking to the population and another linking (with  a nice arrow) to the case-control sample. And then I might want to make the whole thing smaller just a tidbit 

\foreach \x in {11.4, 12, 12.6, 13.2, 13.8, 14.4, 15}
\foreach \y in {-0.1, -0.9, -1.7}
\draw (\x,\y) circle (0.1cm);

\foreach \x in {11.4, 12, 12.6, 13.2, 13.8, 14.4, 15}
\foreach \y in {-0.1, -.9, -1.7}
\draw (\x,\y - 0.1) --  (\x-0.15, \y - 0.4 )-- (\x+0.15, \y - 0.4) -- cycle; 

% cases in row 1

\foreach \x in {13.2}
\filldraw[fill=black!50!white, draw=black] (\x, -.1) circle (0.1cm);

\foreach \x in {13.2}
\filldraw[fill=black!50!white, draw=black] (\x, -.1-0.1) --  (\x -0.15, -.1-0.4)-- (\x+0.15, -.1 - 0.4) -- cycle; 

% cases in row 2
\foreach \x in {11.4, 12, 12.6, 13.2, 13.8}
\filldraw[fill=black!50!white, draw=black] (\x, -.9) circle (0.1cm);

\foreach \x in { 11.4, 12, 12.6, 13.2, 13.8}
\filldraw[fill=black!50!white, draw=black] (\x, -.9-0.1) --  (\x -0.15, -.9-0.4)-- (\x+0.15, -.9 - 0.4) -- cycle; 

% cases in row 3
\foreach \x in {13.8, 14.4}
\filldraw[fill=black!50!white, draw=black] (\x, -1.7) circle (0.1cm);

\foreach \x in { 13.8, 14.4}
\filldraw[fill=black!50!white, draw=black] (\x, -1.7-0.1) --  (\x -0.15, -1.7 -0.4)-- (\x+0.15, -1.7 - 0.4) -- cycle;

% draw the bounding box for the sample
\draw[thick, dashed] (11, -0.7) rectangle (13.5, -2.3);
% add text of the case control sample 

\end{tikzpicture}
    \caption{(a) Illustration of the problem setting: we aim to test conditional null hypotheses $\hnull_j$ in the population $\prob$ while using a selected sample $\law$.  (b) Illustration of a case-control sample. The sample contains equal number of cases and controls while there are fewer cases in the population. }
    \label{fig:collider_bias}
\end{figure}
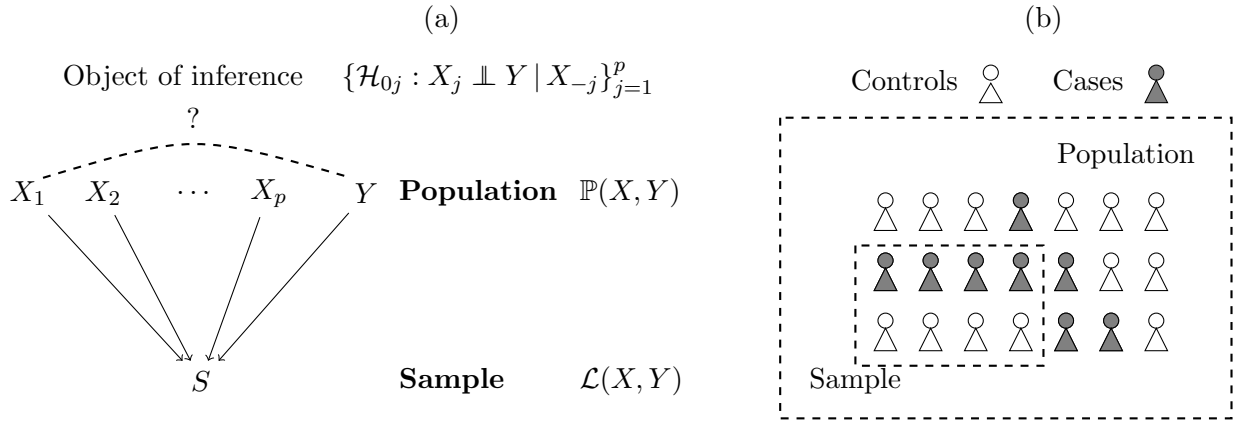

We consider the problem of variable selections, where researchers want to identify among very many variables $X = (X_1,\ldots, X_p)\in\R^p$ those that are related to a response $Y\in\R$ in a population $\prob$ of interest. We want to achieve two objectives simultaneously. First, we want to identify variables that carry unique information on the outcome $Y$, i.e., those variables for which we can reject the conditional null hypothesis 
\begin{equation}\label{eq:cond_null_P}
    H_{0j}: X_{j}\ind Y\,|\,X_{-j}\quad\text{under} \quad \prob. 
\end{equation}
Second, since we are simultaneously testing many hypotheses, we want to control the FDR \cite{benjamini1995} to ensure reproducibility. This is the setting considered in \cite{candes2018modelX}. 
Figure \ref{fig:collider_bias} provides a schematic illustration of how the problem we are interested in departs from this standard framework. The data we get to see is not distributed according to the population distribution $\prob$, which defines the object of inference (\ref{eq:cond_null_P}). Instead, the outcome $Y$ and potential explanatory variables $X=(X_1, \ldots, X_p)$ that we observe have a distribution ${\cal L}(X,Y)$, which reflects the selection mechanism, often corresponding to a case-control design (which is captured in Figure \ref{fig:collider_bias}). Overlooking the fact that $\law\neq \prob$
can lead to the identification of spurious  association or biased estimates. This phenomenon is often referred to as the ``collider bias'', because the variable $S$ representing selection  is a collider in a causal graph and the bias is induced by conditioning on $S$.

To formalize the relation between $\prob$ and $\law$, we consider two situations. Most generally, the sample may be \textit{selected}, where
\begin{equation}\label{eq:selection}
    \law(X, Y) = \prob(X, Y \, | \, S = 1).
\end{equation}
Here, $S = 1$ if the observation is selected and $S=0$ otherwise. If $S\ind Y \,|\, X$,  then a true null hypothesis under $\prob$ is also a null under $\law$. This is not true in general when $S$ depends on both $X$ and $Y$, and thus we are particularly interested in situations when both $X$ and $Y$ affect selection probability. 

In \textit{case-control studies}, as the study in the Paisa region of Colombia and Figure \ref{fig:collider_bias}, the $\law$ is a \textit{mixture} of two  distributions,
\begin{equation*}
    \law(X, Y)  =  p_1 \prob(X, Y | D = 1) + p_0 \prob(X, Y | D = 0),
\end{equation*}
where $D = 1$ ($D=0$) indicates a case (control) respectively, and $p_1$ is the proportion of cases in the sample. %Our goal in both situations is to identify important variables while controlling the FDR. 

Before describing our approach, however, we pause to provide some examples of the challenges we set out to address and a brief overview of recent literature in this topic.

\subsection{Prior work}

Several reviews have documented potential collider bias in empirical studies. In epidemiology, the ``livebirth effect'' refers to the phenomenon that one might observe protective effects of risks on birth defects when studying live born children. For example, one might find maternal smoking to be associated with reduced odds of anencephaly when studying liveborns because both affect survival probability \cite{heinke2020}. In genetic association studies, one may observe a spurious association between one variable (e.g., education) and sex among study participants when both affect participation, and the effect is more pronounced when sex mediates the effect of differentiated participation \cite{pirastu2021sex}. As another example, autosomal genes influencing a trait that is influenced by sex (e.g., height) might be associated with sex when conditional on that trait \cite{Day2016}. Besides these situations that corresponds closely to Figure \ref{fig:collider_bias}, researchers have also explored other causal graph structures that can lead to collider bias, such as when there are  unmeasured confounders \cite{akimova2021ColliderPRS,groenwold2019}. 

Not only can collider bias induce spurious association, it can also lead to biased estimates of the effect of a covariate $X$ on the secondary phenotype $Y$ of interest. For instance, suppose the case-control status $D$ follow a logistic model, then the standard prospective likelihood provide unbiased estimates if and only if $Y$ is independent of $D$ given $X$ \cite{linSPML2009}. In several scenarios, researchers have derived the bias theoretically. For example, Yaghootkar et al. considers a linear model for $Y\,|\, X$, a Probit for $D\,|\, X, Y$, and derived the bias of coefficients when regressing $Y$ onto $X$ using the case-control sample. Groenwold et al. considers several causal graphs and derives the formula for bias when all the variables are Gaussian and follow linear models \cite{groenwold2019, groenwold2021}. 

Besides estimating the bias, researchers have developed several methods to achieve unbiased estimates. For case-control samples, semiparametric maximum likelihood (SPML) fits a full likelihood function for the joint distribution of $(X, Y, D)$. Secondary phenotype regression (SPREG) fits the conditional likelihood conditional on the case-control status $D$ \cite{linSPML2009}. Weighted estimating equations (WEE)   expands the population estimating equation to include $D$. Since only the outcome corresponding to the observed $D$ is observed, they propose to use a counterfactual outcome for the other case \cite{song2016robust}. While these three methods deal with case-control studies, the inverse probability weighting method (IPW) can be applied to general selection mechanism. IPW assigns each observation a weight that is inversely proportional to the selection probability, and one can either optimize the weighted likelihood uses a weighted log-likelihood \cite{ruchardson2007ipw,seaman2013ipw} or solve weighted estimating equation \cite{xing2016robust, Brownstein2022}. Finally, Heckman's method considers a specific case of linear models with bivariate Gaussian errors, and selection is based on truncating one of the variables \cite{heckman1979}. Asymptotic distributions of the estimators are available for all of these methods, and therefore researchers can compute P-values for testing null hypotheses. On the other hand, we expect that the asymptotic theories of these approaches do not apply when the number of variables is large, as statistical inference in the high-dimensional setting is often challenging, and requires assumptions on model sparsity or the covariate distributions e.g., see \cite{highdimrobust_ElKaroui_2013, bellec2024observableadjustmentssingleindexmodels}. 

Compared to estimation, research focusing on hypothesis testing is more limited. We are only aware of \cite{tenzer2022} which tests quasi-independence between two single variables using a weighted permutation test. 

%Suppose there are multiple variables, we aim to test conditional independence $\hnull_j: X_j\ind Y \,|\, X_{-j}$. There are two benefits for testing the conditional null hypothesis. First, it eliminates non-causal variables that happens to be associated with the causal variables. Second, it comes up naturally in a generalized linear model, since testing if a model coefficient is zero is equivalent to \eqref{eq:cond_null_P} (see \cite[Section~3]{candes2018modelX} for more discussions). In addition, we aim to select important variables with replicability guarantee. This work addresses both of these goals, which is formally stated in Section~\ref{section:problem_setting}

\subsection{Summary of contribution and paper outline}
To  test a large number of hypotheses with FDR control, we leverage the model-X knockoff framework \cite{candes2018modelX}. At the core of this procedure is a comparison of  the importance of a variable $X_j$ with its ``knockoff variable'' $\tilde{X}_j$, which is constructed as to be undistinguishable from  $X_j$ given the response $Y$ if $X_j$ is a null variable. Usually,  $(X_i, Y_i)$ are assumed i.i.d.  from the population $\prob$ of interest, resulting in a specific strategy to  sample $\tilde{X}_j$s.   However, when the sample distribution  $\law\neq\prob$,  we  show that this construction no longer guarantees FDR control (\ref{section:model_x_condition}). Inspired by work in \cite{barber2020robustko}, we identify the necessary condition for knockoff filter to achieve its goals when $\law\neq\prob$. Furthermore, we show that we can construct these valid knockoffs by assuming $X$ is from a \textit{tilted} distribution, which tilts the population distribution with the selection probability (Section~\ref{section:tilted_knockoff}). In the remainder of the paper, we discuss how to sample tilted knockoffs in practice and study its FDR and power through simulated examples in Section \ref{section:sampling}.  Section~\ref{section:real} applies the tilted knockoff to the genetic study that motivated this work.

\section{Methods}
\subsection{Model-X knockoff filter when $\prob \neq \law$}\label{section:model_x_condition}

%Lemma \ref{lemma:fdr_control} below summarizes the requirements for effective knockoff  testing of the  
We are interested in testing the null hypotheses $\{{ H}_{0j}\}_{j=1}^p$ (\ref{eq:cond_null_P}) defined with respect to a  population $\prob$ of interest on the basis of a  biased sample with  distribution $\law \neq \prob$.  We use $\hnull$ to  indicate the collection of true null hypotheses.
Before  stating the requirements for a  procedure to work in this setting,  we  introduce the three ingredients of the knockoff framework in the more traditional situation where the sample distribution is the same as the population $\law = \prob$.

 First, for each variable $X_j$, we construct   a synthetic negative control---a  {\em knockoff}---$\tilde{X}_j$ that has no information on $Y$ and mimics the dependence structure among the covariates given $Y$ so closely that, if $X_j$ is null, it is impossible to distinguish it from its knockoff $\tilde{X}_j$. To be precise, a valid knockoff $\tilde{X}_j$ satisfies the exchangeability condition that, 
\begin{equation}\label{eq:condition0}
        (X,\tilde{X})_{\text{swap}(X_j,\tilde{X}_j)}  \stackrel{d}{=} (X,\tilde{X})  \;\; \text{with} \;\; X\sim\prob(X |Y)\quad \forall j \in \hnull.\end{equation}
      Here, $(X,\tilde{X})_{\text{swap}(X_j,\tilde{X}_j)}$ refers to swapping the entries $X_j$ and $\tilde{X}_j$ in $(X,\tilde{X})$. For instance, when $j=2$ and $p = 3$, Eqn.~\ref{eq:condition0} states that 
      $$
      (X_1, X_2, X_3, \tilde{X}_1, \tilde{X}_2, \tilde{X}_3) \stackrel{d}{=}(X_1, \tilde{X}_2, X_3, \tilde{X}_1, X_2, \tilde{X}_3)\;\; \text{with} \;\; X\sim\prob(X |Y). 
      $$
In other words, the second variable cannot be distinguished from its knockoff by using the response $Y$ if it is a null, and the same is true for all the null variables. In \cite{candes2018modelX}, this is achieved by constructing a vector of knockoffs $\tilde{X}$ (a) without looking at $Y$ and  (b) assuring that the following condition holds:
 \begin{equation}\label{eq:condition_modelx}
     (X,\tilde{X})_{\text{swap}(X_j,\tilde{X}_j)} \stackrel{d}{=}(X,\tilde{X})  \;\; \text{with} \;\;  X\sim\prob(X) \quad \forall j,
\end{equation}
where difference between \eqref{eq:condition_modelx} and \eqref{eq:condition0} lies in the distribution of $X$ ($\prob (\cdot)$ vs $\prob (\cdot | Y)$) and on the range of $j$ for which the equivalence is valid (all $j$ vs $j\in \hnull$).

The second step is to construct a test statistics by comparing  the importance of each variable to that of its knockoff.  To do this, we first compute a feature importance statistics $Z_j$ and $\tilde{Z}_j$ for each variable $X_j$ and its knockoff $\tilde{X}_j$. The feature importance statistics are required to be symmetric, i.e., swapping $X_j$ with $\tilde{X}_j$ in the model implies swapping $Z_j$ and $\tilde{Z}_j$ as feature importance. One can typically obtain $Z_j$ and $\tilde{Z}_j$   by fitting a model using both the original variables and their knockoffs, as long as these  have symmetric roles, and use summaries of importance. Examples include the absolute value of LASSO coefficients and Shapley values from a fully connected neural network. Next, one calculates a knockoff score $W_j = f(Z_j,\tilde{Z}_j)$ for each variable by comparing $Z_j$ with $\tilde{Z}_j$ using an anti-symmetric function $f$, such that  $f(Z_j,\tilde{Z}_j) = - f(\tilde{Z}_j, Z_j)$. Combined with the symmetric property of $Z_j$'s, swapping $X_j$ with $\tilde{X}_j$ flips the sign of $W_j$. A large positive $W_j$ indicates that a variable $X_j$ is more informative than its knockoff.  

As a result of the exchangeability condition in Eqn.~\eqref{eq:condition1}, $W_j$ for null variables are equally likely to be positive or negative, and thus one can use the knockoffs to estimate the proportion of false discoveries and determine a threshold that controls the FDR. This is what happens in the third step, when  the threshold $\tau_+$ is computed by
\begin{equation}\label{eq:tau}
        \tau_+ = \min\left\{t>0:\frac{1+\#\{j: W_j\leq -t\}}{\#\{j:W_j\geq t\}}\leq q\right\},
\end{equation}
and an hypothesis is rejected  if its knockoff score $W_j$ exceeds $\tau_+$ . 

Putting these ingredients together we obtain the following lemma, which strongly resembles \cite[Lemma~3.2]{candes2018modelX}.
\begin{lemma}\label{lemma:fdr_control}
    Consider testing the conditional null hypotheses $\{H_{0j}\}_{j=1}^p$ \eqref{eq:cond_null_P} using a sample consisting of $n$ independent observations $(X_i,Y_i){\sim}\law$. If
    \begin{enumerate}[a)]
        \item the knockoff variables $\tilde{X}$ satisfy 
        \begin{equation}\label{eq:condition1}
        (X,\tilde{X})_{\text{swap}(X_j,\tilde{X}_j)}\stackrel{d}{=} (X,\tilde{X}) \;\; \text{with} \;\; X\sim \law(X\,|\, Y)\quad \quad \forall j \in \hnull \;\; (\text{i.e.} \;\; \forall j :  X_{j}\ind_{\prob} Y\,|\,X_{-j}); 
    \end{equation}
    \item the feature importance statistics $(Z_j, \tilde{Z}_j)$ are symmetric w.r.t. $(X_j, \tilde{X}_j)$ and the knockoff score $W_j=f(Z_j,\tilde{Z}_j)$ is defined from an anti-symmetric function $f$;
    \item one selects variables whose knockoff score exceeds the threshold $\tau_+$ defined at level $q$ (\ref{eq:tau});
    \end{enumerate}
    then the false discovery rate of $\hnull_{j}$ is controlled at level $q$.
\end{lemma}

The proof of Lemma~\ref{lemma:fdr_control} uses the same arguments as \cite[Lemma~3.2]{candes2018modelX}. In fact, conditions {\em b.} and {\em c.} are the same, and the distinction lies in  condition {\em a.} Lemma~\ref{lemma:fdr_control} is also similar to \cite[Lemma~1]{gimenez2019multipleKO} where they consider the joint distribution of $(X, \tilde{X}, Y)$ while we consider the conditional distribution $(X, \tilde{X})\,|\, Y$. When the sample distribution $\law$ is different from the population one $\prob$ used to define the hypotheses of interest, a construction as \eqref{eq:condition_modelx} does not guarantee that the $j$-th null variable cannot be distinguished from its knockoff by using the response $Y$. 
 To bring home this point home, we run a simulation.
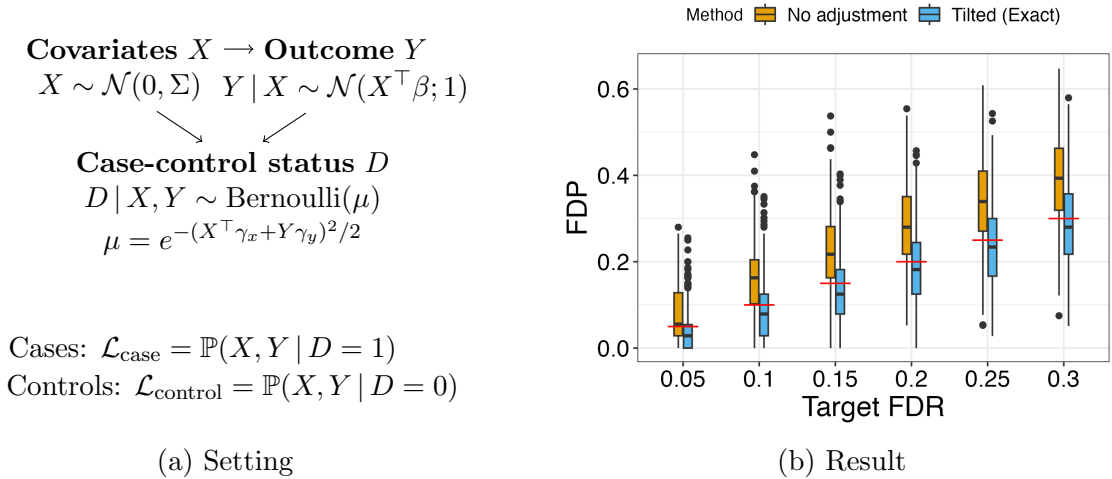
\begin{figure}[ht]
   \begin{center}

    \vspace*{.5cm}
    
    \begin{tikzpicture}
% the populatin diagram
% add a bounding box outside of the population
% \draw[thick, dashed] (-1.5, 0.5) rectangle (4.5, -2.8);

% \draw[thick, dashed] (-1.5, -3.2) rectangle (4.5, -5);

\node[](covariate1) at (0,0) {\textbf{Covariates} $X$};
\node[](covariate2) at (0, -0.5) {$X\sim\N(0,\Sigma)$}; 
\node[](outcome1) at (3,0) {\textbf{Outcome} $Y$};
\node[](outcome2) at (3, -0.5) {$Y\,|\,X \sim \N(X^\top \beta; 1)$};
\node[](casecontrol1) at (1.5,-1.5) {\textbf{Case-control status} $D$};
\node[](casecontrol2) at (1.5,-2) {$D\,|\, X, Y \sim \mathrm{Bernoulli}(\mu)$};
\node[](casecontrol3) at (1.5,-2.5) {$\mu = e^{-(X^\top \gamma_x +  Y \gamma_y)^2/2}$};

\draw [->] (covariate1) -- (outcome1);
\draw [->] (covariate2) -- (casecontrol1);
\draw [->] (outcome2) -- (casecontrol1);

% \node[](population) at (1.5, -3.8) {Population $\prob$};

% the case-control sample diagram

\node[](cases) at (1.1,-4) {Cases: $\law_{\text{case}} = \prob(X, Y\,|\, D = 1)$};
\node[](cases) at (1.5,-4.5) {Controls: $\law_{\text{control}} = \prob(X, Y\,|\, D = 0)$};
% \node[](Casecontrol) at (1.5, -6) {Case-control sample $\law$}
\node[](settings) at (1.4, -5.5) {(a) Setting};

\node[](settings) at (9.6, -5.5) {(b) Result};
\node[anchor=north west](result) at (5.7, 1) {\includegraphics[width = 0.45\textwidth]{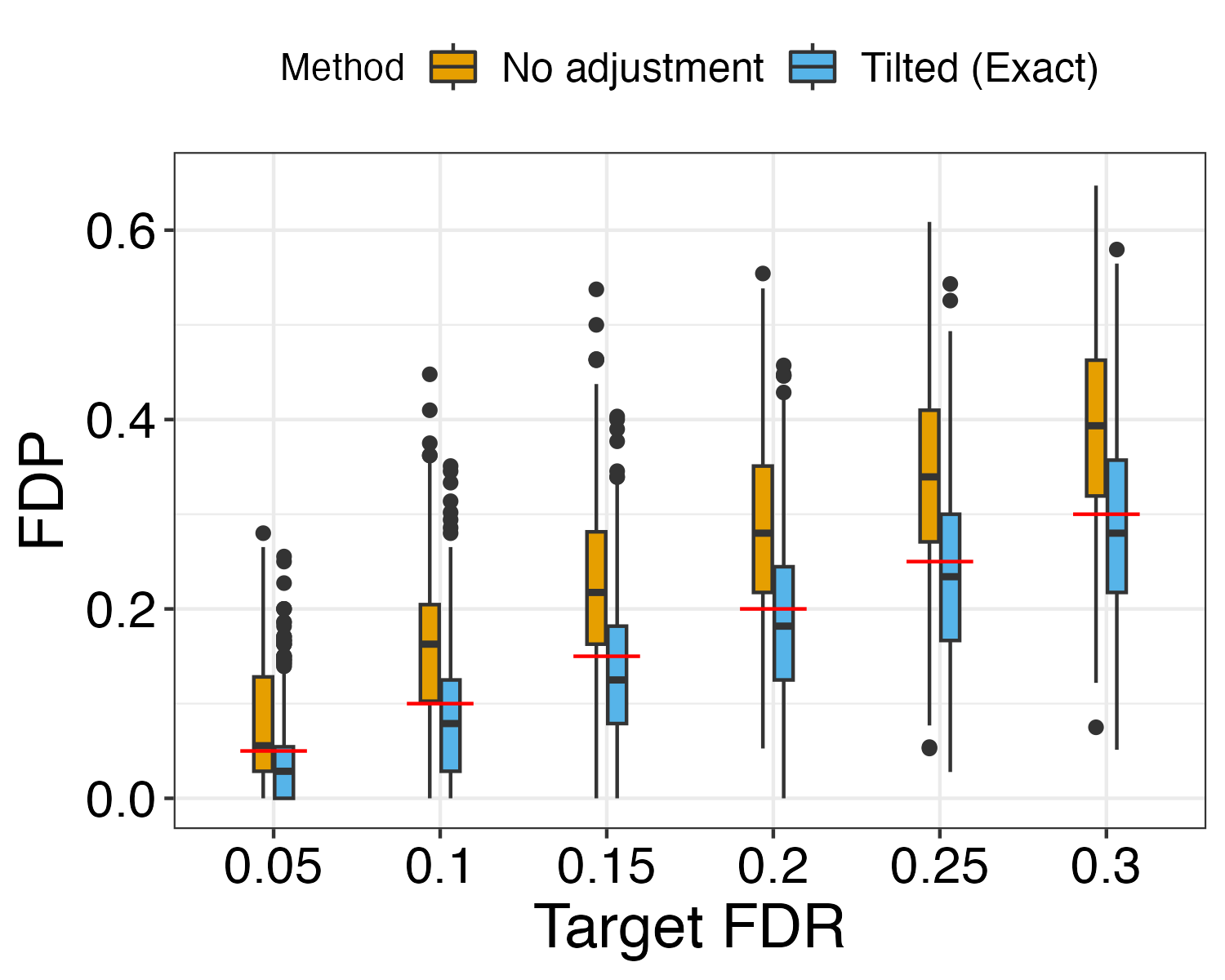}};

\end{tikzpicture}
\end{center} 

    \caption{Simulation result of exact tilted knockoff in Section~\ref{section:tilted_knockoff_defn}: (a) Illustration of the data generating mechanism: first, a large number of $(X_i, Y_i)$ are sampled from the population distribution; then, a case-control status $D$ is sampled based on $(X, Y)$; finally, a fixed number of cases and controls are randomly sampled to form the case-control sample (details of the parameter are reported in Appendix \ref{appendix:selection_detail}). (b) Boxplot of the false discovery proportion (FDP) in $B = 500$ repetitions at each target FDR level. The target FDR levels are also shown in red segments. The yellow boxplot shows the FDP of standard model-X knockoffs without adjusting for sampling bias (``No adjustment''); the blue boxplot shows the FDP of the exact tilted knockoffs (``Tilted (Exact)''). }
    \label{fig:sim}
\end{figure}

We work with a case-control sample, and our response is a secondary phenotype, distinct from the case-control status $D$. In detail, $Y\,|\, X$ is from a linear model, and $D\,|\, X, Y$ is from a Bernoulli distribution (see Figure~\ref{fig:sim} (a)). Importantly, both the response $Y$ and the covariates $X$ affect the probability of being a case. The case-control sample consists of 2000 cases and 2000 controls. 

Figure~\ref{fig:sim} (b) reports  the empirical false discovery proportion (FDP) using the  knockoff  constructed  to satisfy \eqref{eq:condition_modelx} (yellow boxplots). The median FDP is higher than the target FDR level, indicating an inflation in FDR. In comparison, the median FDP is below the target level if  there is no selection, e.g., if we sample $4,000$ observations i.i.d. from the population (see Appendix~\ref{appendix:noselection}, Figure~\ref{fig:no_selection}). 
%This example shows that sampling knockoffs according to \eqref{eq:condition_modelx} may not control the FDR when the sample distribution differs from the population; in Section \ref{section:tilted_knockoff}, we introduce a modification of Eqn.~\eqref{eq:condition_modelx} by tilting $\prob(X)$ with the selection probability that leads to FDR control (Figure~\ref{fig:fdr_gaussian}, green). 

The discrepancy between the population distribution $\prob$, on which we define the target of interest, and the sample distribution $\law$ makes it impossible for the relevant exchangeability condition \eqref{eq:condition1} to be satisfied with a strategy for knockoff sampling that is based only on the marginal distribution of $X$. An additional challenge is  
that $\law(X\,|\, Y)$ is unknown---in sharp  contrast to $\prob(X)$ which is at the center of the property in \eqref{eq:condition_modelx} and assumed known in model X knockoff approaches.
 The main contribution of our work is to show that even if $\law(X\,|\, Y)$ is unknown, it is possible to obtain valid knockoffs, starting from a tilted version of $\prob(X)$.

\subsection{Constructing valid knockoffs for $X$ with a distribution only partially known}\label{section:tilted_knockoff}
%Having shown that the standard model-X knockoff may not control the FDR when we study the secondary phenotype using a case-control sample, we now describe how to define knockoffs that satisfies the sufficient condition in Eqn.~\eqref{eq:condition1}. We formulate an equivalent condition (Section~\ref{section:simplified_condition}) and introduce the tilted knockoffs that satisfies it. We illustrate this idea through two examples: a case-control study (Section~\ref{section:casecontrol}) and conditional randomization test (Section~\ref{section:conditional_randomization}). 
%
%\subsection{What are valid knockoff variables?}\label{section:simplified_condition}

%As noted in Section~\ref{section:model_x_condition}, we cannot directly use Eqn.~\eqref{eq:condition1} to define knockoffs when $\prob\neq \law$ because $\law(X|Y)$ is unknown. 

To make progress, it is useful to write the exchangeability   \eqref{eq:condition1} into a more explicit  form. The joint law of $(X,\tilde{X})$ in   \eqref{eq:condition1} can be represented  as the product of the distribution of  $X$,  $\law(X\,|\, Y)$,  which is {\em specified} by the problem we are interested in,   and  the distribution of the knockoffs $\tilde{X}$ given $X$,  $\tilde{Q}_y(\tilde{X}\,|\, X),$ which is a matter of  {\em user's choice}. 
%
%Let $\tilde{Q}_{y}(\tilde{X}\,|\, X)$ denote the distribution of the knockoffs $\tilde{X}$ given $X$, wherein the subscript $y$ indicates that the $\tilde{Q}$ might depend on values of the response $Y$. Given $Y = y$, the joint law of $(X,\tilde{X})$ is   
%\begin{equation*}
%    \law_y(X,\tilde{X}) := \law(X,\tilde{X}\,|\, Y = y) = \law(X\,|\, Y)\tilde{Q}_y(\tilde{X}\,|\, X). 
%\end{equation*}
For a given value $Y=y$, then, the pairwise exchangeable condition \eqref{eq:condition1} can be thus written as 
\begin{equation}\label{eq:condition1_Q}
 \law(X_j,X_{-j}\,| Y\! = \!y) \tilde{Q}_y(\tilde{X}_j, \tilde{X}_{-j} | X_j, X_{-j}) \!=\! \law(\tilde{X}_j,X_{-j}| Y\! =\! y) \tilde{Q}_y({X}_j, \tilde{X}_{-j}| \tilde{X}_j, X_{-j}).
\end{equation} 

Given any distribution for $X$ there are now several methods have been developed to construct a distribution  $\tilde{Q} (\tilde{X}|X)$ that would satisfy the exchangeability condition  for all $j$ \cite{bates2021metro,romano2020deep,spector2022KOpower,gimenez2019multipleKO}, and indeed, the selection of $\tilde{Q}$ that might  yield high power has also been studied, see \cite{gimenez2019multipleKO, spector2022KOpower}. Here we will not consider this problem, and simply assume that for a given a distribution for $X$, we have a   knockoff generating mechanism $\tilde{Q}$ that satisfies the pairwise exchangeable condition.
%, i. e.
%\begin{equation}\label{eq:exchangeability_q}
%    Q(X_j,X_{-j})\tilde{Q}(\tilde{X}_j, \tilde{X}_{-j}\,|\,X_j,X_{-j}) = Q(\tilde{X}_j,X_{-j})\tilde{Q}({X}_j, \tilde{X}_{-j}\,|\,\tilde{X}_j,X_{-j}),\quad \forall j\in\hnull. 
%\end{equation}

We focus, instead, on the fact that the distribution of $X$, $\law(X\,|\, Y = y) $,  is in our problem unknown, and investigate strategies to approximate it with a distribution $Q_y(X)$ that can be plugged into the knockoff machinery,  retaining FDR control. In doing so, we take inspiration from the work of  \cite{barber2020robustko}, which studies robustness of the model-X knockoff when the distribution of $X$ is estimated and the knockoffs is pairwise exchangeable w.r.t. that estimated distribution.   Lemma \ref{lemma:condition_simple} is our first step.

\begin{lemma}\label{lemma:condition_simple}
Suppose $(X, Y) \sim \law(X, Y)$. Let 
\begin{enumerate}[a)]
\item the distribution $Q_y(X)$ satisfy
    \begin{equation}\label{eq:conditional}
    Q_{y}(X_{j}\,|\,X_{-j}) = \law(X_{j}\,|\,X_{-j}, Y = y) \quad \forall j\in\hnull \quad \text{and} \quad \forall y;
    \end{equation}
 \item   $\tilde{Q}_y(\tilde{X}\,|\,X)$ satisfy the exchangeability condition w.r.t. $Q_y$:
\begin{equation}\label{eq:exchangeability_q}
    Q_{y}(X_j,X_{-j})\tilde{Q}_{y}(\tilde{X}_j, \tilde{X}_{-j}\,|\,X_j,X_{-j}) = Q_{y}(\tilde{X}_j,X_{-j})\tilde{Q}_{y}({X}_j, \tilde{X}_{-j}\,|\,\tilde{X}_j,X_{-j}),\quad \forall j\in\hnull.
\end{equation}
\end{enumerate}
 Then $(X,\tilde{X})$ satisfies Eqn.~\eqref{eq:condition1}. 
\end{lemma}

Lemma~\ref{lemma:condition_simple} follows from  \cite[Lemma~1]{barber2020robustko} and a short proof is provided in Appendix~\ref{appendix:proof_kl}.   
%\cite{barber2020robustko} studies robustness of the model-X knockoff when the $\prob(X)$ is estimated and the knockoffs is pairwise exchangeable w.r.t. an estimated distribution $\hat{\prob}(X)$.  \cite{barber2020robustko} shows that the upper bound of the FDR inflation depends on the KL divergence between conditional distributions $\hat{\prob}(X_j\,|\, X_{-j})$ and $\prob(X_j\,|\, X_{-j})$. It turns out that this insight also allows us to characterize appropriate $Q_y$ when $\law(X\,|\, Y)$ is unknown.  
Note that using Lemma~\ref{lemma:condition_simple}, we can easily verify that the ``standard'' model-X knockoff \eqref{eq:condition_modelx} controls the FDR when $(X_i, Y_i)$ are i.i.d. samples from a population $\prob$, since $\law(X_j\,|\, X_{-j}, Y) = \prob(X_j\,|\, X_{-j})$ for null variables, and thus setting $Q_y = P$ satisfies Eqn.~\eqref{eq:conditional}. 

In the next section, we show how to arrive at a suitable  $Q_y$, by ``tilting'' $\prob$.

% Considering $\tilde{Q}_y$ from this family provides flexibility while still being practical, because approximating $\tilde{Q}_y$ for a known distribution $Q_y$ is a well studied problem and several software are available for this task \cite{bates2021metro,romano2020deep}. 

\subsection{The tilted knockoffs}\label{section:tilted_knockoff_defn}

 The results we have obtained so far, indicate a strategy based on the following steps: 
 \begin{enumerate}[a)]
    \item Choose a probability model $Q_{y}(X)$ with  conditional distributions of $X_{j}$ given $X_{-j}$ identical to  $\law(X_j\,|\, X_{-j}, Y)$ for all $j$ (see Eqn.~\eqref{eq:conditional}).  
    \item Choose a mechanism $\tilde{Q}_y(\tilde{X}\,|\, X)$ to generate $\tilde{X}$ that satisfies the pairwise exchangeable condition with respect to $Q_y$ (see Eqn.~\ref{eq:exchangeability_q}). 
    \item Generate knockoff variables $\tilde{X}$ using $\tilde{Q}_y(\tilde{X}\,|\, X)$. 
\end{enumerate}
How to identify a $Q_y(X)$ with the needed properties? The following theorem shows that if we are working with a selected sample, we can {tilt} the marginal distribution $\prob(X)$ by the selection probability $\prob(S=1\,|\, X, Y)$. A short proof is provided in Appendix~\ref{appendix:proof_main}.

\begin{theorem}\label{thm:main}
Suppose $\law(X, Y) = \prob(X, Y\,|\, S = 1)$, then the distribution 
\begin{equation}\label{eq:qy_main}
    Q_y(X) \propto \prob(X) \prob(S = 1\,|\, X, Y = y).
\end{equation}
satisfies
\begin{equation*}
   Q_y(X_j\,|\,X_{-j}) = \law(X_j\,|\,X_{-j}, Y)
\end{equation*}
for each null variable $j\in\hnull$. 
\end{theorem} 

We now apply this strategy to the simulation illustrated in Figure~\ref{fig:sim}. In this example, $Q_y(X)$ is a mixture of two Gaussians (see Appendix~\ref{appendix:Qy_simulation}); thus, we apply the knockoff sampling method for a mixture of Gaussians proposed in \cite{gimenez2019multipleKO}. We observe in Figure \ref{fig:sim} that the tilted knockoff (blue) controls the FDR, as expected. 

While illustrative of the potential of tilted knockoffs,  this example is somewhat contrived: on the one hand, we design the joint distribution so that we know how to sample exact knockoffs, while in reality this might be difficult; on the other hand, we assume that we know the selection probability, while in practice this would often need to be estimated. We will relax both these assumptions in Section \ref{section:sampling}. In the remainder of this section, we consider two applications of the tilted knockoff strategy: studying a secondary phenotype in a case-control study and conditional randomization tests. 

\subsection{Tilted knockoffs and case-control studies}\label{section:casecontrol}

We now turn to our motivating example, where researchers collected a case-control sample, and ask the question: can we use knockoffs to analyze both primary and secondary phenotypes? For the primary phenotype, the problem has already been considered and we know that the knockoff continues to control the FDR \cite{barber2019knockoff_casecontrol}. The framework we introduced provides an easy way to prove this: in this case,  $\law(X\,|\, Y) = \prob(X\,|\, Y, D) = \prob(X\,|\, Y)$ since the outcome of interest $Y$ is the case-control status $D$. Choosing $\prob(X)$ as $Q_y(x)$ thus satisfies Eqn.~\eqref{eq:conditional} in Lemma~\ref{lemma:condition_simple} by definition of the conditional null hypotheses. We now turn to the problem of studying secondary phenotypes and define tilted knockoffs that control the FDR. 

\begin{proposition}\label{prop:case_control}
Consider a case-control sample where we know the case-control status $D$ of each individual and study a secondary phenotype $Y$. Define a tilted distribution $Q_{y,d}(X)$ that depends on the phenotype $Y$ and the case-control status $D$ as 
\begin{equation}\label{eq:qy_casecontrol}
    Q_{y, d}(X) \propto \prob(X) \prob(D = d\,|\, X, Y = y), 
\end{equation}
and sample knockoffs according to $\tilde{Q}_{y, d}(\tilde{X}\,|\, X)$ which is pairwise exchangeable w.r.t. $Q_{y,d}(X)$, then the knockoff+ procedure controls the FDR. 
\end{proposition}
 
Since $\law(X, Y) = \prob(X, Y\,|\, D)$, Proposition~\ref{prop:case_control} follows from verifying  $Q_{y,d}(X)$ satisfies Eqn.~\eqref{eq:conditional}. We show in simulations (see Figure~\ref{fig:estimated_logistic_main} and details in Appendix~\ref{appendix:so_estimated}) that the tilted knockoff controls the FDR whereas the FDR of the standard knockoff can be highly inflated. 

% Finally, we briefly comment on the situation of a matched pair design, where researchers match a control with a case sample by similarity of their covariates. We can think of such as design as first sampling a case $(X_1, Y_1)\sim\prob(X, Y\,|\, D=1)$ and then a control sample from a \textit{reweighted} distribution according to its similarity with the case $(X_2,Y_2)\,|\, (X_1,Y_1)\sim\prob(X_2, Y_2\,|\, D=0)w(X_1, Y_1)$. Using a similar reasoning, we can show that the tilted distribution for $(X_1, X_2)$ is proportional to 
% \begin{equation*}
%     \prob(X_1)\prob(X_2)w(X_1,X_2)\prob(D=1\,|\, X_1, Y_1)\prob(D=0\,|\, X_2, Y_2). 
% \end{equation*}
% Here, we consider the two observations simultaneously because of their correlation. The exchangeability condition for the knockoff generating mechanism $\tilde{Q}_{y1, y2}(\tilde{X}_1,\tilde{X}_2\,|\, X_1,X_2)$ now becomes one can swap both $(X_{1,j},X_{2,j})$ and their knockoffs $(\tilde{X}_{1,j},\tilde{X}_{2,j})$ without changing the joint distribution. \zq{can delete this paragraph. If $Y = D$ the standard knockoff still might not control the FDR in a matched pair design unless we can swap $(X_{1,j}, X_{2, j})$ with $(\tilde{X}_{1,j},\tilde{X}_{2,j})$ without changing the joint distribution of $(X_1, X_2,\tilde{X}_1,\tilde{X}_2)$.}

\subsection{Tilted knockoffs and the conditional randomization test}\label{section:conditional_randomization}

In this section we apply the idea of tilted knockoffs to conditional randomization tests based on biased samples. The conditional randomization test (CRT) provides valid P-values for any test statistics $T$ for testing an individual conditional independence hypothesis in Eqn.~\eqref{eq:cond_null_P} \cite[Supplement~F]{candes2018modelX}. Briefly, suppose we  observe $n$ pairs $(X_i, Y_i)\stackrel{iid}{\sim} \prob$. If $X_j$ is a null variable, then the conditional distribution of $X_j$ given $X_{-j}$ and $Y$ satisfies $X_j \,|\, X_{-j}, Y\stackrel{d}{=} X_j \,|\, X_{-j}$. To compute the P-value of CRT, we resample the $j$-th variable of each observation $X_{i,j}^{(k)}\sim\prob(X_{i,j} \,|\,X_{i,-j})$ and compute the test statistics $T_j^{(k)}:=T(X_{j}^{(k)},X_{-j}, Y)$. Repeating this process $K$ times yields $K$ test statistics. Under the null hypothesis, conditional on $(X_{-j},Y)$, they are independent and identically distributed as the observed statistics $T_j := T_j(X_j,X_{-j}, Y)$. Therefore, a one-sided P-value can be defined as
\begin{equation}\label{eq:crt_pvalue}
    P_j = \frac{1}{K+1}\left[1+\sum_{k=1}^K \mathrm{I}(T_j^{(k)}\geq T_j)\right], 
\end{equation}
and it is stochastically larger than a uniform variable \cite{candes2018modelX}.  

Suppose we work with a biased sample where the sample distribution $\law$ differs from the population  $\prob$ of interest, then the same challenges encountered by knockoffs might be encountered by CRT and a similar solution applies. By our arguments before,  it suffices to sample $X_j\,|\,X_{-j}$ from any distribution whose conditional distribution of $X_j\,|\, X_{-j}$ equals $\law(X_j\,|\,X_{-j}, Y)$ as shown in the next Proposition. In particular, $Q_y(X)$ defined in Eqn.~\eqref{eq:qy_main} satisfies this requirement. 

\begin{proposition}\label{prop:crt}
    Suppose $(X_i, Y_i)\stackrel{iid}{\sim}\law$, and  we test the conditional null hypothesis $X_j\ind Y\,|\, X_{-j}$ under $\prob$. If we resample $X^{(k)}_j \sim \law(X_j\,|\,X_{-j}, Y)$, then the P-value defined in Eqn.~\eqref{eq:crt_pvalue} are stochastically larger than uniform variables. 
\end{proposition}
\begin{proof}
Under $\hnull$, the resampled observations satisfy
\begin{equation}
    X_j^{(k)} \,|\, X_{-j}, Y \stackrel{d}{=} X_j \,|\,X_{-j}, Y.
\end{equation}
Therefore, conditional on $(X_{-j}, Y)$, $T_j^{(k)}$ are independent and $T_j^{(k)} \stackrel{d}{=} T_j$. 
\end{proof}

\section{Implementation strategies}\label{section:sampling}

We have shown theoretically how to generate knockoff variables for a selected sample: (1) define a tilted distribution $Q_y(X)$ with tilts the population distribution by the selection probability; (2) generate tilted knockoffs according to $\tilde{Q}_y(\tilde{X}\,|\, X)$ that is pairwise exchangeable w.r.t. ${Q}_y$. Two questions come up when we apply this theory in practice: first, how to choose $\tilde{Q}_y$? second, does the tilted knockoff still control the FDR when we estimate the selection probability? We consider these two questions in this section.

\subsection{Second-order tilted knockoff}\label{section:second_order_tilted}

Supposing for now that we know tilted distribution $Q_y(X)$, and thus our task is to sample knockoffs from a distribution $\tilde{Q}_y(\tilde{X}\,|\, X)$ that is pairwise exchangeable w.r.t. $Q_y(X)$. Several methods have been developed for this task: the metropolized knockoff \cite{bates2021metro} can generate exact knockoffs efficiently if the covariates are from a graphical model with moderate tree width. On the other hand, it is computationally intensive if the tree width of $Q_y(x)$ is large, which happens for example when $X$ is from a Markov chain but the selection probability depends on many variables. The deep knockoff method \cite{romano2020deep} generates knockoffs using a neural network and thus can sample knockoffs for any unknown distribution, as long as we have sufficient training samples. Since $Q_y(X)$ depends on the response $Y$, we would need to train a new neural network for every different $Y$, and this can be computationally challenging. We show examples using the metropolized and deep knockoffs in  Appendices~\ref{appendix:deep} and \ref{appendix:metro}, but mindful of the challenges above, we resort to a strategy based on second order approximation.

The ``second order'' knockoffs are introduced in  \cite{candes2018modelX}. The idea is to limit exchangeability requirement to the first two moments. This of course results in exact knockoffs when the distribution of $X$ is Gaussian. In our setting,  we can think about this strategy as approximating $Q_y(X)$ by a Gaussian distribution $\N(\mu_y,\Sigma_y)$ matching the the mean $\mu_y$ and the covariance $\Sigma_y$ of $Q_y(X)$ and then  choosing $\tilde{Q}_y(\tilde{X}\,|\,X)$ to be pairwise exchangeable w.r.t. the Gaussian distribution.

To estimate the two moments of $Q_y(X)$ defined in Eqn.~ \eqref{eq:qy_main}, we suggest using importance sampling \cite{bds3rd}. This strategy is feasible under two conditions. First, we need to be able to  generate i.i.d. samples from the marginal distribution $\prob(X)$. This is a reasonable assumption in most of the cases where we would imagine to use knockoffs, and certainly for genetic data. Second, we need to  know, or can  estimate, the selection probability (see next section). In detail, we sample $K$ i.i.d. samples $X_k\stackrel{iid}{\sim}\prob(X)$, and then we estimate $\E_{Q_y}[h(X)]$ of a statistics $h(x)$ by 
\begin{equation*}
    \widehat{E}_{Q_y}[h(X)] = \frac{\frac{1}{K}\sum_{k=1}^K h(X_k)w_y(X_k)}{\frac{1}{K}\sum_{k=1}^K w_y(X_k)},
\end{equation*}
where $w_y(X_k) = \prob(S = 1\,|\, X_k, Y = y)$. We choose $ h_1(x) = x$ to estimate the mean $\hat{\mu}_y$, and $h_2(x) = xx^\top$ estimate the second moment $\hat{\mu}_y^{(2)}$. We estimate the covariance by $\hat{\Sigma}_y = \hat{\mu}_y^{(2)} -  \hat{\mu}_y\hat{\mu}_y^\top$. 

\begin{figure}[t]

\centering
     \begin{tikzpicture}
% the populatin diagram
% add a bounding box outside of the population
% \draw[thick, dashed] (-1.5, 0.5) rectangle (4.5, -2.8);

% \draw[thick, dashed] (-1.5, -3.2) rectangle (4.5, -5);

\node[](covariate1) at (0,0) {\textbf{Covariates} $X$};
\node[](covariate2) at (0, -0.5) {$X\sim\N(0,\Sigma)$}; 
\node[](outcome1) at (3.5,0) {\textbf{Outcome} $Y$};
\node[](outcome2) at (3.5, -0.5) {$Y\,|\,X \sim \N(X^\top \beta; 1)$};

\node[](casecontrol1) at (1.5,-2.5) {\textbf{Selection} $S$};
\node[](casecontrol2) at (1.5,-3) {$S\,|\, X, Y \sim \mathrm{Bernoulli}(\sigma(X, Y))$};
\node[](casecontrol3) at (1.5,-3.5) {$\sigma(X, Y) = 1 / (1+e^{ - X^\top \gamma_x -\gamma_y Y - \gamma_0})$};

\draw[->] (covariate1) -- (outcome1);
\draw [->] (covariate2) -- (casecontrol1);
\draw [->] (outcome2) -- (casecontrol1);

\node[anchor=north west](result) at (5.7, 1.1) {\includegraphics[width = 0.45\textwidth]{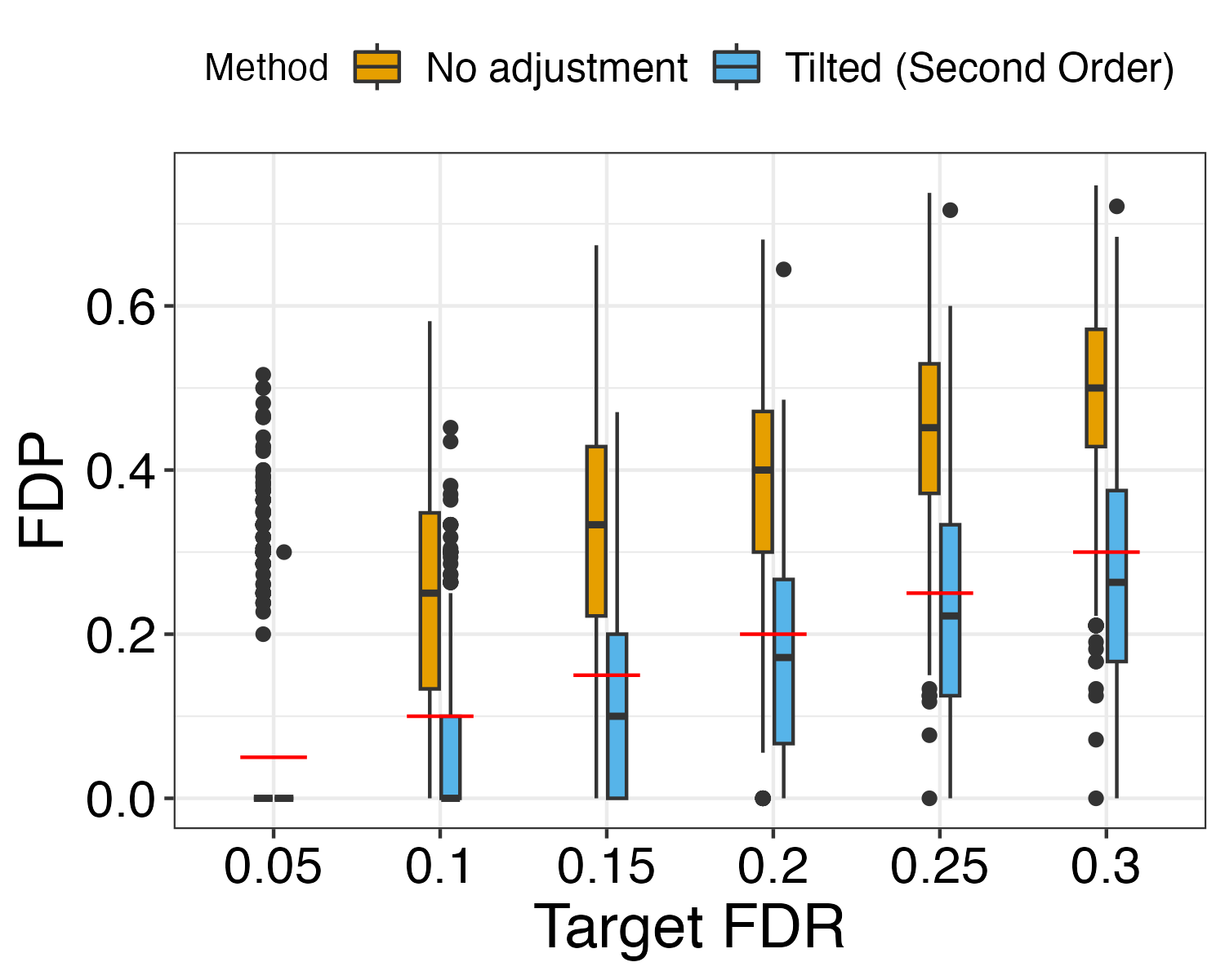}};

% add A and B somewhere 
\node[](fig1) at (1.5, -5.4) {(a) Setting};
\node[](fig2) at (9.8, -5.4) {(b) Result};
\end{tikzpicture}

    \caption{Simulation result of the second order knockoff in Section~\ref{section:second_order_tilted}: (a) Illustration of the data generating mechanisms; details of the parameters are reported in Appendix \ref{appendix:so_detail}. (b) Boxplot of the FDP in $B = 522$ repetitions at each target level. The yellow boxplot shows the FDP of standard model-X knockoffs without adjusting for sampling bias (``No adjustment''); the blue boxplot shows the FDP of the second order knockoffs (``Tilted (Second Order)'').}
    \label{fig:second_order}
\end{figure}

We illustrate the second order tilted knockoff with a simulation, where the selection probability is set by  a logistic model, and  no closed-form expression is available  to sample the tilted knockoffs (see Appendix~\ref{appendix:so_detail} for details).  The second order tilted knockoff controls the FDR while the FDR for standard model-X knockoff is heavily inflated (Figure~\ref{fig:second_order}). We also observe that the standard knockoff generally makes more selections so that a higher FDR is accompanied by higher power.  (see Figure~\ref{fig:power_second_order} in Appendix~\ref{appendix:so_detail}). 

\subsection{Estimating parameters in the selection probability}\label{section:so_estimated}

The selection probability $P(S = 1\,|\, X, Y)$ is typically unknown and needs to be estimated from data. While strategies to do this are available, it is important to evaluate the impact on FDR of working with an estimate rather than the true value of $P(S = 1\,|\, X, Y)$. We do this via a simulation study.

We generate a case-control sample, where the number of cases and number of controls are both equal to $n = 2000$ and the number of variables is $p = 200$. The covariates are sampled from a Markov chain, and both the response and the case-control status are sampled from a logistic model. We report the simulation details in Appendix~\ref{appendix:so_estimated}.

\begin{figure}[t]
    \centering
    \includegraphics[width = 1\textwidth]{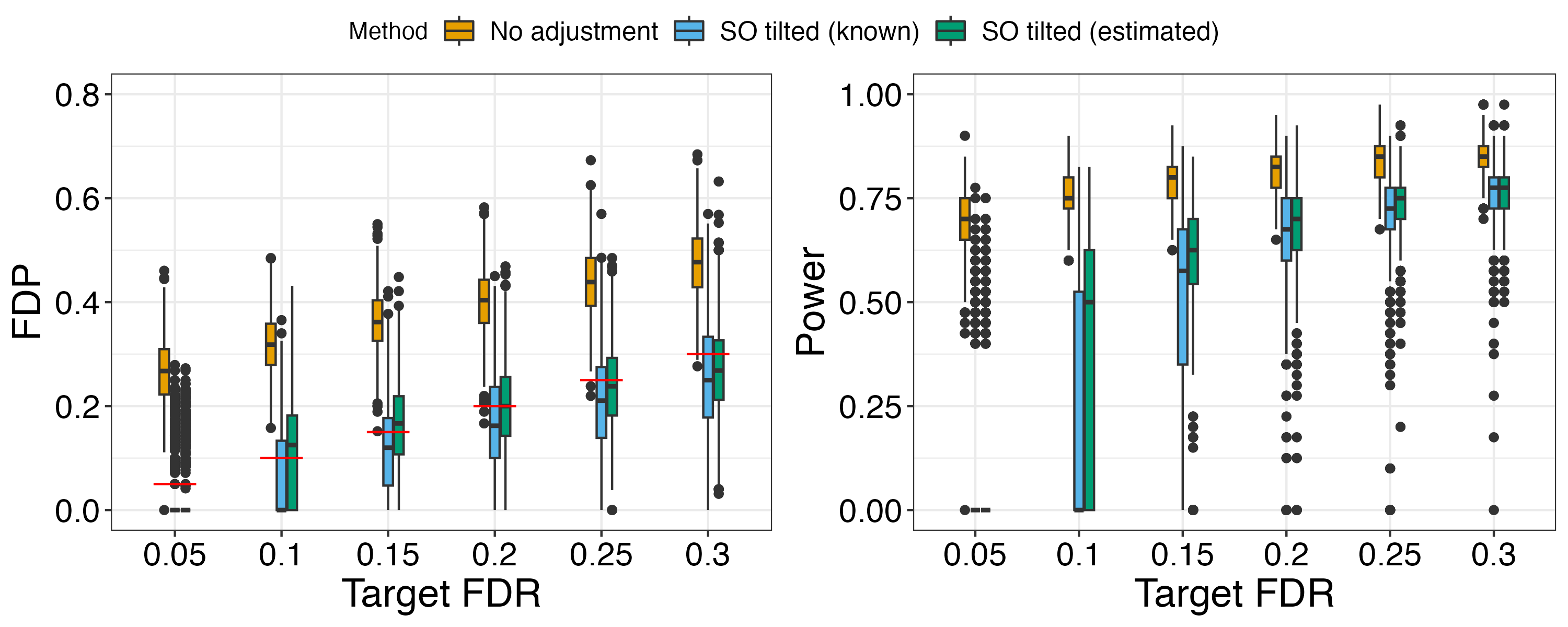}
    \caption{Boxplot of the FDP and power of second order tilted knockoffs with estimated parameters in $B=2,000$ repetitions. We simulate a case-control sample and the response is a secondary phenotype. The standard knockoff without adjustment (``No adjustment'') does not control the FDR, whereas the second order knockoff using known (``SO tilted (known)'') or estimated parameters both control the FDR (``SO tilted (estimated)''). Standard knockoff without adjustment has the highest power among the three methods.}
    \label{fig:estimated_logistic_main}
\end{figure}

We apply three approaches:  (1) standard knockoff without adjustment, second order tilted knockoff using (2) known or (3) estimated selection probability. We estimate the selection probability using a logistic regression and adjusting  the intercept to match  the population prevalence of the cases (see \cite[Section~4.3]{mccullagh1989}). We report the FDR and power using each method in Figure~\ref{fig:estimated_logistic_main}. The FDR of standard knockoff without adjustment (``No adjustment'') nearly doubles the target level. In comparison, the second order tilted knockoff using either known (``SO tilted (known)'') or estimated parameters (``SO tilted (estimated)'') controls the FDR, and the power of these two approaches are comparable (Figure~\ref{fig:estimated_logistic_main}). 

As observed before (Figure~\ref{fig:power_second_order}, the tilted knockoff is slightly less powerful than the standard model-X knockoff. For instance, at target FDR level $q = 0.3$, the standard knockoff is able to discover approximately 35 of the non-nulls on average, and it makes about 32 false discoveries. On the other hand, the tilted knockoff discovers about 30 of the non-nulls, and it makes only about 12 false discovery on average. This power vs FDR tradeoff is reasonable.
%Therefore, if a variable selected by both the standard and the tilted knockoff to be a non-null variable, we would have higher confidence that they are truely non-nulls and can assign higher priority when designing follow-up studies. 

We consider two more methods to estimate parameters in a sparse logistic model, an $\ell_1$ regularized logistic and a two-step knockoff selection procedure. We observe that both methods control the FDR and have similar power to the logistic regression in this example (see  Figure~\ref{fig:logistic_estimated_appendix} in Appendix~\ref{appendix:so_estimated}).

To generalize these results, one might wonder the effect of model misspecification for the selection probability. In general, one can calculate a theoretical upper bound of the FDR inflation when $Q_y(X)$ is incorrectly estimated using the same argument in \cite{barber2020robustko}. This theoretical upper bound tends to be conservative, because it bounds the FDR when one uses the sharpest statistics to distinguish  a variable from its knockoff. Instead of quantitative calculations, we make a few observations: first, even if the form of the model is incorrect (e.g., we fit a logistic model when the true model is a probit), as long as the estimated probability $\prob(D = 1\,|\, X, Y)$ is close to the true value, we expect that the tilted knockoff would approximately control the FDR. This means that one can use a flexible method with high prediction accuracy to obtain this probability. Second, even if the estimate of $\prob(D = 1\,|\, X, Y)$ is incorrect, suppose $Q_y(X_j\,|\, X_{-j})$ is correctly estimated, then by the same argument as before, we expect the tilted knockoff to still approximately control the FDR. 
 
\section{Secondary phenotypes in a SMI case-control study}\label{section:real}

% Maybe reference to PGC? 

Having shown how to implement the tilted knockoff, we turn to our motivating question in Section~1:  can we identify genetic variants associated with a secondary phenotype in a case-control study for severe mental illness?    

The data we analyze is relative to individuals from the Paisa region in Colombia and is collected as a case-control sample for severe mental illness  (see \cite{service2020} for details of the study design). We exclude one individual from pairs that are related at a level higher than first degree cousin, to be able to assume that  our data is i.i.d. given the diagnosis $D$. The sample size of diagnosis $D$ is reported in Table~\ref{tab:sample_size}. In this study, individual genotypes, demographic information, medication use are collected, diagnosis and psychiatric symptomatology are obtained by structured interview \cite{maruish_assessment_symptom}, and neurocognitive performances are assessed by nine cognitive tasks from the Penn Computerized Neurocognitive Battery \cite{cnb_Gur_2010}. For the purpose of illustrating how to apply the tilted knockoff, we focus on genotype data $X$, diagnoses $D$ and neurocognitive performances $Y$. 

\begin{table}[h]
    \centering
    \begin{tabular}{|c| c | l|c|}
    \hline
    & Total & Diagnosis    &  Sample size\\
    \hline 
  \multirow{4}{*}{Case}  &  \multirow{4}{*}{3961} & Schizophrenia    &  604 \\
  &  &  Major depressive disorder & 1785 \\
  &  &  Bipolar disorder I & 1173 \\
  &  & Bipolar disorder II & 399 \\
  \hline   
Control  &     1158 & - & 1158\\
    \hline
    \end{tabular}
    \caption{Number of individuals who have completed the Penn Computerized Neurocognitive Battery (CNB) in the sample.}
    \label{tab:sample_size}
\end{table}

While the tilted knockoff method can be applied to all of the secondary phenotypes, we focus on two neurocognitive domains that are hypothesized to be associated with severe mental illnesses, highly heritable, and show strong differences by diagnoses: attention, which is assessed by the Penn Continuous Performance test, and processing speed, which is assessed by Digit Symbol Test. Both accuracy  (number of correct responses) and speed (median response time for correct items) are measured in each test. This yields four secondary phenotypes $Y$: attention (speed/accuracy) and processing speed (speed/accuracy). We carry out individual analysis of each of these variables. For simplicity, we use attention (speed) as an example in our exposition, which we denote as $\atts$, and note that the analysis is repeated for each of the four secondary phenotypes. As $\atts$ is a continuous variable and estimating one tilted distribution for each value is computationally intensive, we discretize discretize the measurements using  the top quintile as a cut-off point, and analyze the binary response $\atts\in\{0, 1\}$. 

We start with 704 genetic variants that have been identified in external studies to be associated with at least one of the diagnosis in other populations \cite{gwas_bipolar_mullins2021, Trubetskoy_2022, depression_Als_2023}. This allows us to explore if their effect is reproducible in the Paisa population and if they are associated to the secondary phenotypes. After filtering variants to eliminate pairwise correlations above 0.8, we obtain $p = 623$ candidate variants.  We aim to test $\hnull_{j}: X_j\ind \atts\,|\, X_{-j}$ in the Paisa population for genetic variants $X_j$. 
  
We  define a tilted distribution $Q_{y,d}(x)$ for each value of diagnosis $d$ and secondary phenotype $y\in\{0, 1\}$ as 
\begin{equation}\label{eq:tilted_distribution_real}
    Q_{y,d}(x)\propto \prob(x) \prob(D = d\,|\, X=x, \atts = y).
\end{equation}
(see Proposition~\ref{prop:case_control}). In Eqn~\eqref{eq:tilted_distribution_real}, $\prob(x)$ refers to the population distribution of genotypes $X$ and $\prob(D = d\,|\, X=x, \atts=y)$ is the probability of diagnosis $d$ given genotype $x$ and secondary phenotype value $y$. 

We now estimate individual terms in Eqn.~\eqref{eq:tilted_distribution_real}. We conveniently approximate the population distribution $\prob(X)$ by a Gaussian. This might appear rather arbitrary for genetic variants, but it should be understood simply as focusing on the first two moments, and it has been proved a successful strategy \cite{chu_groupKO_2024}. Since we do not have access to a random sample from the Paisa population, we use a weighted estimate from our case-control sample. We calculate the mean $\hat{\mu}_d$ and covariance matrices $\hat{\Sigma}_d$ using data from individuals with each diagnosis $d$ and estimate the population mean $\hat{\mu}$ and covariance $\hat{\Sigma}$ by the weighted average using population incidence $w_d$ of each diagnosis  reported in \cite{Geospatial_Song_2022}: 
\begin{equation}
\hat{\mu} = \sum w_d \hat{\mu}_d,\quad \text{and}\quad \hat{\Sigma} = \sum w_d \hat{\Sigma}_d. 
\end{equation}
This approximated population distribution is also used to sample standard knockoﬀs.

We then use a two step approach to estimate the probability of each diagnosis. First, we  fit an $\ell_1$ regularized logistic regression using the diagnosis $D \in\{d, \text{control}\}$ as response and $X$ and $\atts$ as covariates. This leads us to  a list of selected variables for each diagnosis, which we combine with variants known to be associated with the diagnosis to form a set of relevant covariates $\mathcal{S}_{d}$. Finally, we fit a logistic regression to estimate the selection probability $\prob(D\,|\, X, \atts)$ using genetic variants in $\mathcal{S}_{d}$ and $\atts$ as covariates. We further adjust the estimated intercept since the sample proportion differs from the population incidence.  We report the pseudo-$R^2$ of the logistic regression in Appendix Table~\ref{tab:pseudoR2}. The pseudo-$R^2$ is  highest for predicting schizophrenia, typically around 30\%, and lowest for predicting bipolar disorder for $Y = \text{processing speed}$, at only 1-2\%. We plot the linear predictors in Appendix Figure~\ref{fig:att_a_linear_predictor}, which shows that higher neurocognitive functions are associated with lower probability of each diagnosis. 

After we obtain the tilted distribution $Q_{y,d}(x)$, we approximate its mean and covariance 
$(\tilde{\mu}_{y,d},\tilde{\Sigma}_{y,d})$ for each diagnosis $d$ and $y\in\{0,1\}$ using importance sampling. We then generate second order tilted knockoff accordingly. To obtain more stable selections, we use multiple knockoffs of size 3 as proposed in \cite{gimenez2019multipleKO}.  We repeated the knockoff generation 100 times, and thus we obtain 100 sets of selected variables in total for each secondary phenotypes. We report the details of knockoff generation in the Appendix~\ref{app:real_knockoff_generation}. 

\begin{figure}[t]
    \centering
    \includegraphics[width = 0.8\textwidth]{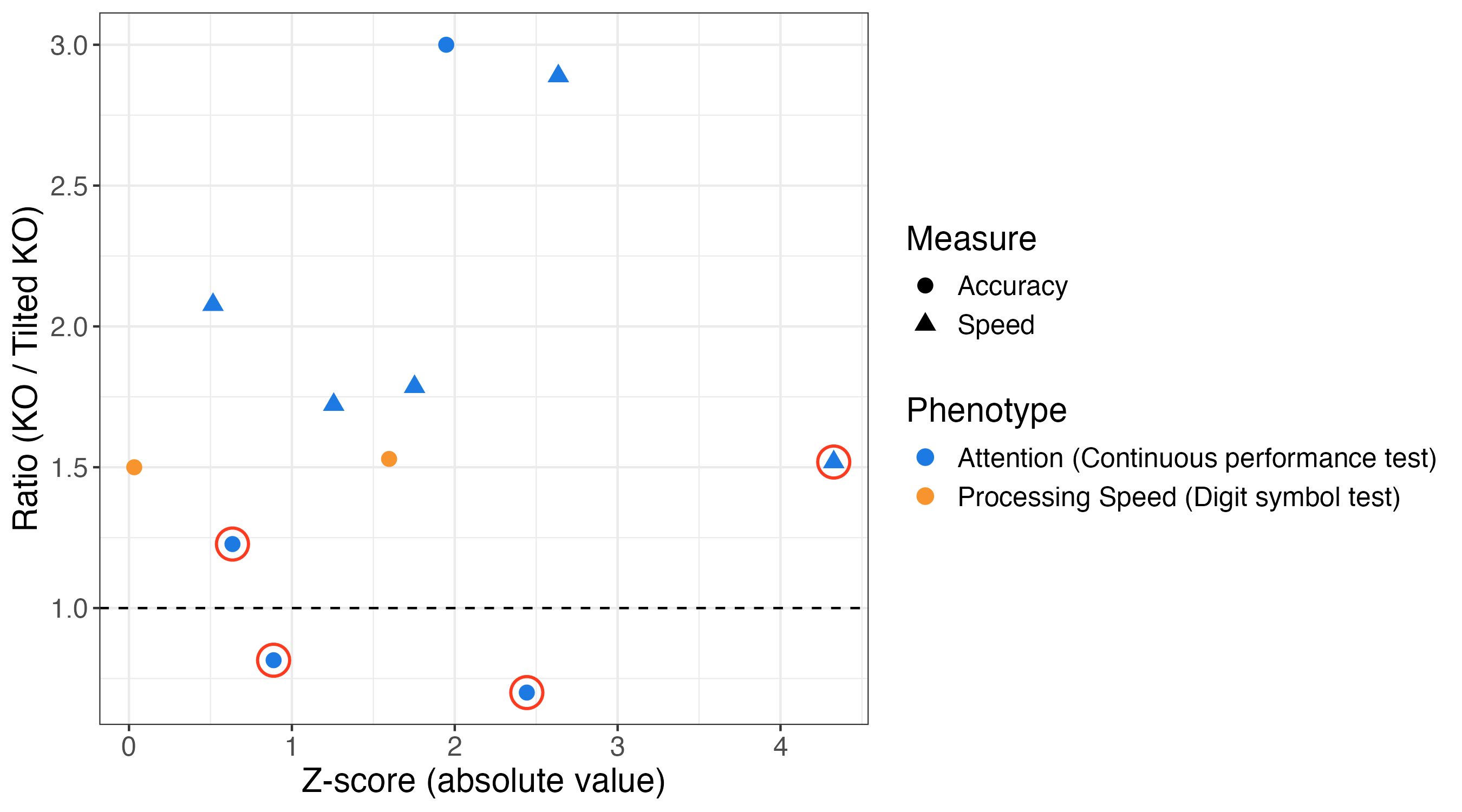}
    \caption{Ratio of times a genetic variant is selected by standard knockoff versus tilted knockoff method. The $z$-score is obtained by fitting a probability-weighted logistic regression. This figure shows 11 out of 623 variants which are part of at least 20 of the 100 selection sets identified by different runs of the knockoff procedures; variants selected at least 20 times by tilted knockoffs are highlighted by red circles. Different neurocognitive domains are indicated by color and different measures by shapes. Points above the dashed horizontal line are selected more frequently by standard knockoff compared to tilted knockoff.}
    \label{fig:nselect_real}
\end{figure}

We compute the proportion of times a genetic variant is selected by standard knockoff and tilted knockoff out of the 100 repetitions and display the variables selected at least 15\% of times for any trait in Figure~\ref{fig:nselect_real}. The standard knockoff typically selects a variable more frequently compared to the tilted knockoff, which is expected as we observed in simulations that standard knockoff is more powerful; at the same time, we know that it does not control the FDR so that its selections are not trustworthy. As ground truth is not available in this example, we attempt to validate the selections using inverse probability weighted analysis, where each observation is weighted by the inverse of $\prob(D\,|\, X, \atts) / \pi_d$, using all the variables selected by either standard or or tilted knockoff over 20\% of times. This approach is similar to the robust secondary phenotype analysis proposed in \cite{xing2016robust}; here, we divide the estimated probability of each diagnosis by the population incidence to obtain more balanced weights. The standard deviation is calculated by the standard error in 1,000 nonparametric bootstrap with fixed weights. The absolute values of $z$-scores are shown on the horizontal axis in Figure~\ref{fig:nselect_real}, and higher values may indicate higher likelihood that a variant is a non-null. Variants with z-scores less than 2 are more frequently selected by standard knockoff compared to tilted knockoff, and this shows that tilted knockoff may make fewer false discoveries compared to standard knockoff. We note that the neither approach selects any variant over 50\% of the times in this study, these findings should be considered as exploratory and more data is needed to robustly identify genetic variants associated with the secondary phenotypes.

% \zq{explain the variable selected at the highest time -- this is important as this is the scientific problem}

% \zq{add the total number of secondary phenotypes and add that one can analyze a lot more. number of prior associated SNPs}

% \zq{check what the estimated probability is given the genotype and phenotype, histogram of probability, as phenotype and genotype changes (in the supplement), equivalent of $R^2$ in the main text, helps to understand the significance of the problem -- explain how this affects how important the tilting is}

% \zq{move some details knockoffs in the supplement (e.g., MVR versus ME knockoffs)}

% \zq{how many are selected for multiple times -- explain that we can trust this threshold, if we cannot get more than 50\% then we can say that we are not able to find really robust solution right now, these are suggestive and we are collecting more data -- explain that we don't have enough data to answer the problem -- not much power, can explain the variant with the highest selection times -- take this as an exploratory findings}

% \begin{table}[h]
%     \centering
%     \begin{tabular}{|c|l|c c|}
%     \hline
%     & Diagnosis    &  \# with CNB & \# without CNB \\
%     \hline 
%   \multirow{5}{c}{Case}  & Schizophrenia    &  604& 529\\
%   &   Major depressive disorder & 1785 & 853\\
%   &   Bipolar disorder I & 1173 & 1101\\
%   &   Bipolar disorder II & 399 & 146\\
%   \hline 
  
%   &   Control & 1158 & 184\\
%     \hline
%     \end{tabular}
%     \caption{Number of individuals with and without Computerized Neurocognitive Battery (CNB) measurements in the data.}
%     \label{tab:sample_size}
% \end{table}

\section{Discussion}

This paper concerns knockoff variable selections in the analysis of a secondary phenotype $Y$ when the sample is not representative of the population of interest. We show that, in this context, the key property knockoffs should satisfy is the pairwise exchangeability condition \textit{conditional on $Y$}. This observation motivates our definition of the tilted knockoffs, and may be important in other applications of the knockoff method across multiple populations.  

We show in simulated studies that standard knockoff methods without accounting for sampling bias might lead to highly inflated FDR, while the tilted knockoff controls the FDR. We applied the tilted knockoff procedure to identify genetic variants related to neurocognitive performance related to severe mental illnesses in a case-control sample, demonstrating how to apply tilted knockoff in real world settings. 

While we demonstrated how to implement the tilted knockoff, generating tilted knockoffs is still computationally intensive. For instance, in order to use the second order tilted knockoff, we need to estimate the mean and covariance of the tilted distribution for every potential value of the response $Y$, and this is computationally challenging when $Y$ is a continuous variable. One question for future research is exploring whether the second order approximation of tilted distribution can be approximated by a discrete set of values of $Y$. 

In simulations, we observed that the FDR is controlled when the model is correctly specified and the sample size is relatively large, but it is unclear to what extent an incorrectly specified model would affect the FDR. Quantifying how incorrectly estimated selection probability would affect the chance of a null variable to be selected could offer valuable information for practitioners to interpret selections and decide which discoveries to prioritize.

\appendix
\renewcommand{\thefigure}{A\arabic{figure}}
\setcounter{figure}{0}

\section{Simulation details}

\subsection{Exact tilted knockoff}\label{appendix:selection_detail}

We start by sampling $N = 40,000$ i.i.d. observations $(X_i \in\R^p, Y_i \in\R, D_i\in\{0,1\})$, $p = 400$)  from a population $\prob$ defined as following: the covariates $X\sim\N(0,\Sigma)$ where $\Sigma$ is a block diagonal matrix with block size $B = 10$; within each block, $\Sigma_{ij} = 0.5^{|i-j|}$. The response $Y$ is from a linear model $Y = X^\top \beta + \varepsilon$, where $\varepsilon\sim\N(0, 1)$. 10\% of the $\beta_j$ are sampled to be non-nulls, and magnitude of non-nulls are  $\beta_j\sim\N(0,0.5^2)$. The case-control status $D_i$ is sampled according to 
\begin{equation}
\prob(D_i = 1\,|\, X_i, Y_i) = e^{-v_i^2/2}, \quad\text{where}\quad v_i =X_i^\top \gamma_x +  Y_i \gamma_y.  
    \end{equation}
20\% of $\gamma_{x,j}$ are sampled to be non-nulls and the non-null $\gamma_{x,j}$ are i.i.d. $\gamma_{x,j}\sim\N(0, 0.5^2)$. The effect size of the response $Y$ on the case-control status is $\gamma_y = 2$.

After sampling $N$ observations from $\prob$, we sample $n_1 = 2,000$ cases ($D= 1$) and $n_0 = 2,000$ controls ($D= 0$) i.i.d. from the cases or controls.

We apply the standard Gaussian model-X knockoff to generate knockoff variables; in particular, we use the approximate SDP knockoff \cite{knockoffR}.  

\begin{figure}[ht]
    \centering
    \includegraphics[width = 0.55\textwidth]{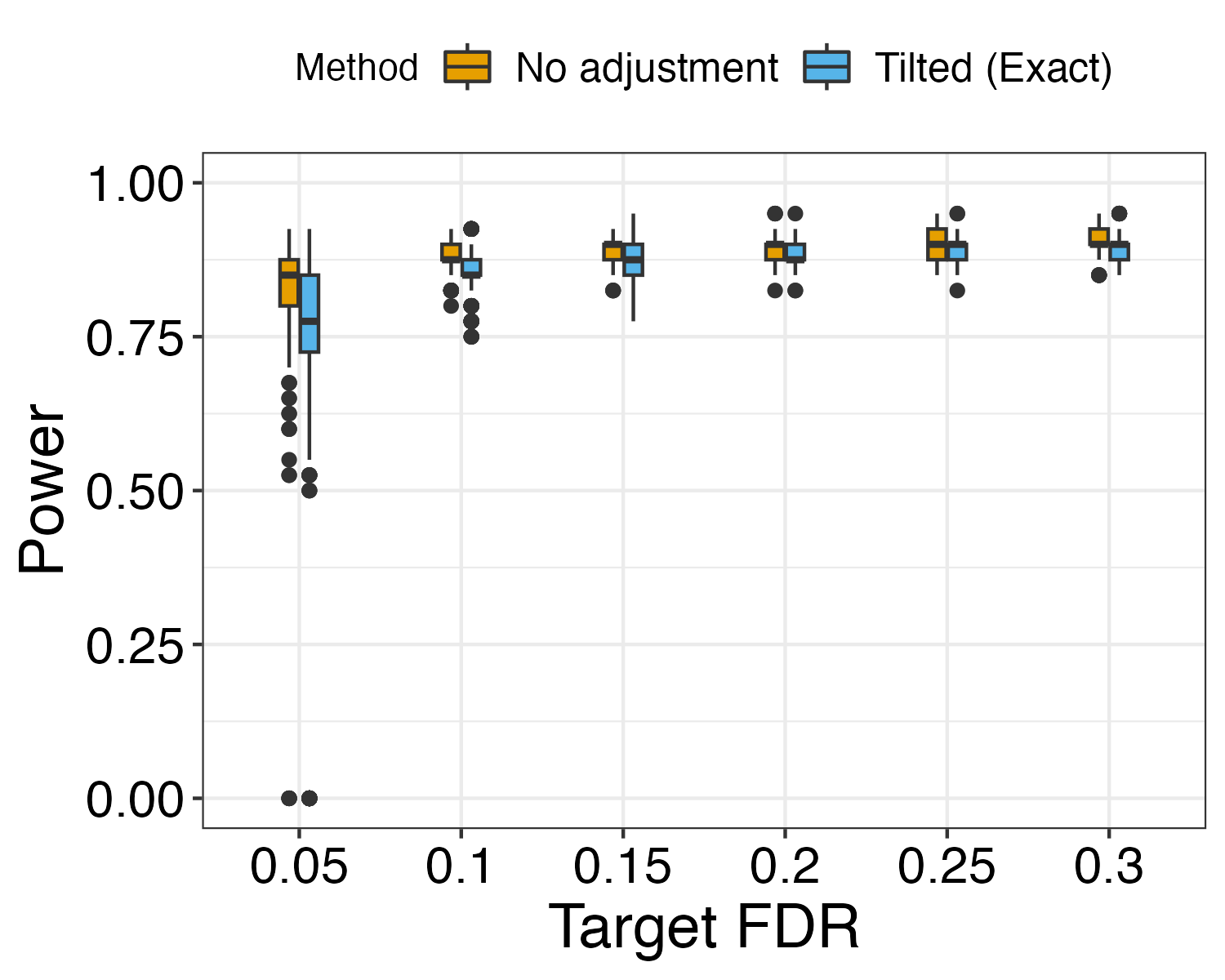}
    \caption{Comparison of power of the standard knockoff and second order tilted knockoff; see Section~\ref{section:model_x_condition}. }
    \label{fig:power_exact}
\end{figure}

We define the knockoff statistics by the \texttt{stat.lasso\_lambdasmax} in  \texttt{knockoff} package \cite{knockoffR}. In detail, we fit a LASSO regression of $Y$ onto  $(X, \tilde{X})$ and define the feature importance statistics for each $X_j$ and $\tilde{X}_j$ as
\begin{equation}
    Z_j = \sup\{\lambda: \hat{\beta}_j(\lambda)\neq 0\}
\end{equation}
where $\lambda$ controls the size of the $L_1$ penalty. We define the knockoff score as 
\begin{equation}
    W_j = (Z_j \vee Z_{j+p})\cdot \sign(Z_j -Z_{j+p}),
\end{equation}
and use the knockoff+ filter to select variables.  We repeat the simulation $B = 500$ times and show a boxplot of false discovery proportion (FDP) in Figure~\ref{fig:second_order} (b) and ratio between  the number of true discoveries and total number of non-null variables (power) in Figure~\ref{fig:power_exact}. 

\subsection{FDR without selection}\label{appendix:noselection}
We simulate an example without selection by sampling $N = 4,000$ i.i.d. observations from the population defined in Appendix~\ref{appendix:selection_detail}. Figure~\ref{fig:no_selection} shows that the standard knockoff method controls the FDR in this setting. 

\begin{figure}[ht]
    \centering
    \includegraphics[width = 0.5\textwidth]{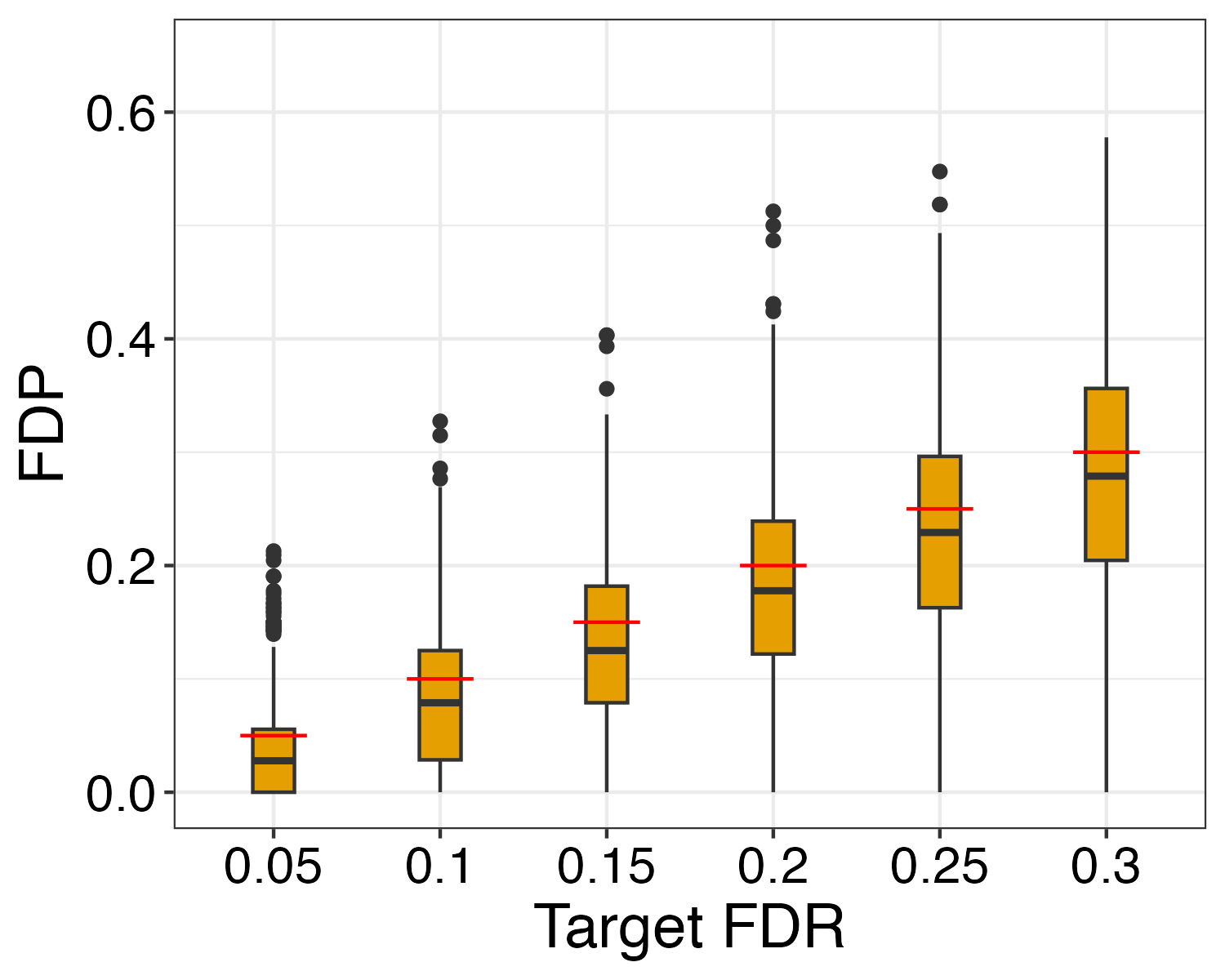}
    \caption{Boxplot of the false discovery proportion (FDP) in $B = 500$ repetitions using standard knockoff when there is no selection. The target FDR are shown in red segments. The setting is the same as in Section~\ref{section:model_x_condition}, except that now each sample consists of $N=4000$ random observations from the population.  }
    \label{fig:no_selection}
\end{figure}

\subsection{Second order tilted knockoffs}\label{appendix:so_detail}

\begin{figure}
    \centering
    \includegraphics[width = 0.55\textwidth]{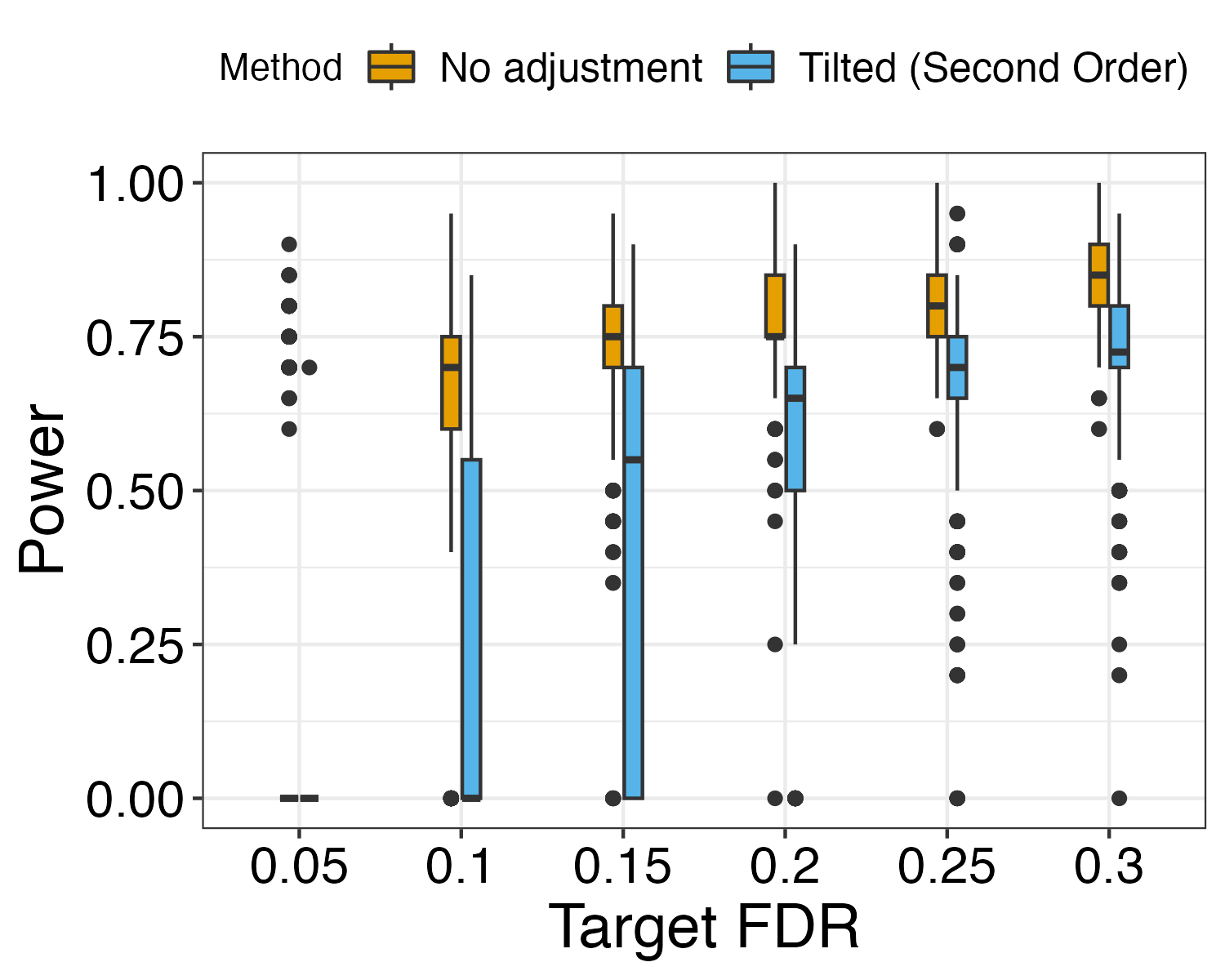}
    \caption{Comparison of power of the standard knockoff and second order tilted knockoff; see Section~\ref{section:second_order_tilted}.}
    \label{fig:power_second_order}
\end{figure}

We start by sampling $N = 5,000$ observations $(X_i \in\R^p, Y_i \in\R, S_i\in\{0,1\}, p = 200)$ i.i.d. from a population $\prob$. The joint distribution $\prob(X, Y)$, and the distribution of $\beta_j$ is the same as in Appendix~\ref{appendix:selection_detail}. The selection probability $S_i$ is sampled according to 
\begin{equation}
\prob(S_i = 1\,|\, X, Y) = \frac{1}{1+e^{-v}}, \quad\text{where}\quad v =  X^\top \gamma_x + Y\gamma_y  +  \gamma_0.   
    \end{equation}
20\% of $\gamma_{x,j}$ are sampled to be non-nulls and the non-null $\gamma_{x,j}$ are i.i.d. $\gamma_{x,j}\sim\N(0, 0.25^2)$, $\gamma_y = 2$, and the intercept is $\gamma_0 = -4$. Our final sample consists of a sub-sample of the population with $S_i = 1$. We use the same knockoff statistics as in Appendix~\ref{appendix:selection_detail}. We repeat the simulation $B = 522$ times and show a boxplot of false discovery proportion (FDP) in Figure~\ref{fig:second_order} (b) and ratio between  the number of true discoveries and total number of non-null variables (power) in Figure~\ref{fig:power_second_order}. 

\subsection{Second order tilted knockoffs with estimated parameters}\label{appendix:so_estimated}
In this section we describe simulation details in Section~\ref{section:so_estimated}.

We sample the covariates $X$ from a three-state Markov chain at equilibrium. The initial probability for each state is equal to $1/3$, and the transition probability matrix
\begin{equation*}
    \prob(X_{j+1} = k\,|\, X_j = l) = P_{l, k},
\end{equation*}
is defined as 
\begin{equation*}
P = \left(\begin{matrix}
    0.5 & 0.3 & 0.2 \\
    0.2 & 0.5 & 0.3 \\
    0.3 & 0.2 & 0.5 
\end{matrix}\right).
\end{equation*}
The matrix $X$ is then centered to have mean equal to 0.

We sample responses from a logistic model 
\begin{equation*}
    \prob(Y = 1\,|\, X) = \frac{1}{1+\exp(-X^\top \beta)},
\end{equation*}
where there are 40 non-null variables and coefficients of the non-nulls are from $\N(0, 0.4^2)$.

Finally, we sample the case-control probability from a logistic model of the form
\begin{equation*}
    \prob(D = 1\,|\, X, Y) = \frac{1}{1+\exp(-\gamma_0 + X^\top \gamma_x + Y \gamma_y)}.
\end{equation*}
Here, $\gamma_0 = -6$, which means that the cases are uncommon, 40 of the $\gamma_{x,j}$ are non-zero and the non-zero $\gamma_{x,j}$ are from $\N(0, 0.4^2)$. The coefficient $\gamma_y = 2$, which means that $Y$ has a relatively strong effect on the case-control status.

\begin{figure}
    \centering
    \includegraphics[width = 1\textwidth]{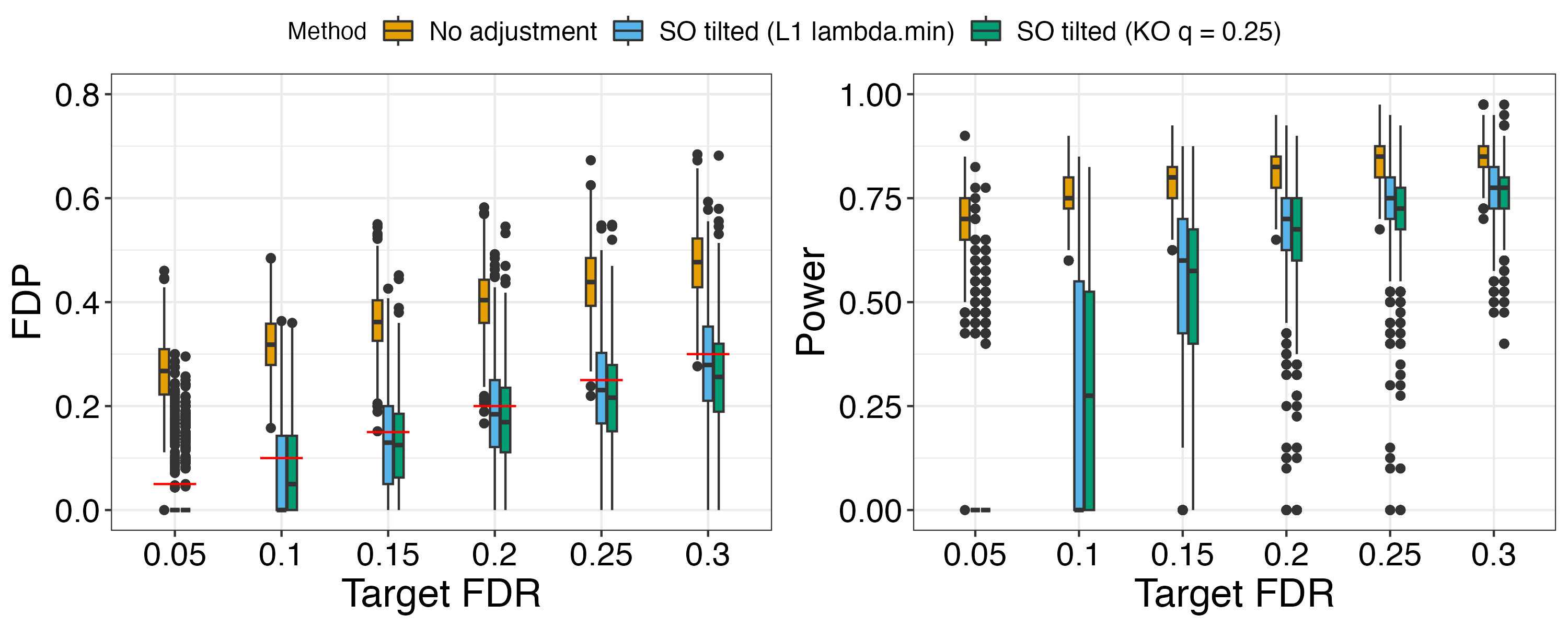}
    \caption{Boxplot of the FDP and power of second order tilted knockoffs with estimated parameters in $B=2000$ repetitions. We simulate a case-control sample and the response is a secondary phenotype. We consider two methods to estimate selection probability here: (a) fitting a $\ell_1$ regularized logistic regression (``SO tilted (L1 lambda.min)'') and (b) a two-stage procedure where standard knockoff at threshold $q = 0.25$ are used to select variables related to case-control status $D$ and a logistic regression is estimated with these variables as covariates and $D$ as response (``SO tilted (KO $q=0.25$)''). }
    \label{fig:logistic_estimated_appendix}
\end{figure}

Figure~\ref{fig:estimated_logistic_main} shows the FDP and power of standard and tilted knockoffs. Here, we consider two additional methods to estimate model parameters and report the FDP and power in Figure~\ref{fig:logistic_estimated_appendix}. (1) We use the $\ell_1$ regularized logistic regression with regularization parameter at the value that minimizes cross-validation error (``SO tilted (L1 lambad.min)''). (2) We use a two-stage procedure. In the first stage, we use standard model-X knockoff to select variables related to the case-control status $D$. In the second stage, we fit a logistic regression model for $D$ using the selected variables as covariates to estimate the probability of being a case (``SO tilted (KO q = 0.25)'').  Both methods are able to control the FDR and they have slightly lower power compared to the standard knockoff. 

\section{Proofs}

\subsection{Proof of Lemma~\ref{lemma:condition_simple}}\label{appendix:proof_kl}
We verify Eqn.~\eqref{eq:condition1_Q} directly and note that our proof is the same as \cite[Lemma 1]{barber2020robustko} except that we condition further on the outcome $Y$. 

Let $X_j$ be a null variable and w.l.o.g. we assume that $X_j$ is discrete. Then it suffices to show that for any possible values of $(X, Y)$, the joint distribution of $(X, \tilde{X})$, where $X\sim\law(X\,|\, Y)$ and $\tilde{X}\,|\, X\sim\tilde{Q}_y(\tilde{X}\,|\, X)$ satisfies the pairwise exchangeable condition for $X_j$ and $\tilde{X}_j$. We focus on the values of the $j$-th variable $X_j$ and its knockoff $\tilde{X}_j$ and denote all the variables other than the $j$-th variable as $X_{-j}$ and similarly for the knockoffs $\tilde{X}_{-j}$. 
\begin{align*}
    & \law(X_j = a, X_{-j} = x_{-j} \,|\, Y = y)\tilde{Q}_y(\tilde{X}_j = b,\tilde{X}_{-j} = \tilde{x}_{-j}\,|\,X_j = a, X_{-j}=x_{-j}) \\
 = & \law(X_{-j} = x_{-j}\,|\, Y = y) \law(X_j = a\,|\, X_{-j} = x_{-j}, Y = y) \tilde{Q}_y(\tilde{X}_j = b,\tilde{X}_{-j} = \tilde{x}_{-j}\,|\,X_j = a, X_{-j}=x_{-j})  \\
= & \law(X_{-j} = x_{-j}\,|\, Y = y) Q_y(X_{j} = a\,|\,X_{-j} = x_{-j}) \tilde{Q}_y(\tilde{X}_j = b,\tilde{X}_{-j} = \tilde{x}_{-j}\,|\,X_j = a, X_{-j}=x_{-j}) \\
= & \law(X_{-j} = x_{-j}\,|\, Y = y) Q_y(X_{j} = b\,|\,X_{-j} = x_{-j}) \tilde{Q}_y(\tilde{X}_j = a,\tilde{X}_{-j} = \tilde{x}_{-j}\,|\,X_j = b, X_{-j}=x_{-j}) \\
= & \law(X_{-j} = x_{-j}\,|\, Y = y) \law(X_{j} = b\,|\,X_{-j} = x_{-j}, Y = y) \tilde{Q}_y(\tilde{X}_j = a,\tilde{X}_{-j} = \tilde{x}_{-j}\,|\,X_j = b, X_{-j}=x_{-j})\\
= & \law(X_j = b, X_{-j} = x_{-j}\,|\, Y = y) \tilde{Q}_y(\tilde{X}_j = a,\tilde{X}_{-j} = \tilde{x}_{-j}\,|\, X_{j} = b,\tilde{X}_{-j}=\tilde{x}_{-j})
\end{align*}

In the third and fifth equality, we use the definition of $Q_y(X)$ in Eqn.~\eqref{eq:conditional}. In the forth equality we use the fact that that $\tilde{Q}_y(\tilde{X}\,|\, X)$ satisfies pairwise exchangeable condition for $X_j$ and $\tilde{X}_j$ w.r.t. $Q_y(x)$ (Eqn.~\eqref{eq:exchangeability_q}).

\subsection{Proof for Theorem~\ref{thm:main}}\label{appendix:proof_main}

We prove the case for discrete variables below, and the continuous case is the same. 
Since  $\law(X, Y) = \prob(X, Y\,|\, S = 1)$, 
\begin{align*}
 &  \law(X_j =   t\,|\, X_{-j} = x_{-j}, Y = y) \\
 = &  \prob(X_j = t\,|\, X_{-j} = x_{-j}, Y = y, S = 1) \\
 = &   \frac{\prob(X_j = t, X_{-j} = x_{-j}, Y = y, S=1)}{\prob(X_{-j} = x_{-j}, Y = y, S=1)}\\
  = & \frac{\prob(X_j=t,X_{-j}=x_{-j})\prob(Y = y\,|\, X_j = t,X_{-j} = x_{-j})\prob(S=1\, |\, X_j = t, X_{-j}=x_{-j}, Y=y)}{\sum_{s}\prob(X_j = s, X_{-j}=x_{-j})\prob(Y=y\,|\, X_j = s,X_{-j}=x_{-j})\prob(S=1\,|\,  X_j = s, X_{-j} = x_{-j}, Y = y)} \\
= &\frac{\prob(X_j = t,X_{-j}=x_{-j})\prob(S=1|X_j = t,X_{-j}=x_{-j}, Y = y)}{\sum_{s}\prob(x_j = s,X_{-j}=x_{-j})\prob(S=1|X_j = t, X_{-j} = x_{-j}, Y = y)}\\
=& Q_y(X_j=t\,|\,X_{-j}=x_{-j}).
\end{align*}
The fourth equality is because $X_j\in\hnull$ under $\prob$, thus $\prob(Y\,|\,X_{-j}, X_j) = \prob(Y\,|\, X_{-j})$. 

\subsection{Derivation of $Q_y$ in Section \ref{section:tilted_knockoff}}\label{appendix:Qy_simulation}
Let $X\in\R^p$ denote the covariates, $Y\in\R$ the response of interest, $D \in\{0, 1\}$ refer to the the case-control status, and $S \in\{0, 1\}$ indicator of selection. Then, by Eqn.~\eqref{eq:qy_main},
\begin{align*}
        Q_y(x) &\propto  \prob(x)\prob(S = 1\,|\, X, Y) \\
        & = \prob(x )\left\{\prob(S=1, D = 1\,|\, X, Y) + \prob(S=1, D = 0\,|\, X, Y) \right\}.
    \end{align*}
Here, 
\begin{align*}
    & \prob(S = 1, D = 1\,|\, X, Y)  \approx \frac{n_1}{\# \{ D = 1\}}\prob(D = 1\,|\, X, Y)=\frac{n_1}{\# \{ D = 1\}} e^{-v^2/2},
\end{align*}
and 
\begin{align*}
    & \prob(S = 1, D = 0\,|\, X, Y)  \approx \frac{n_0}{\# \{D = 0\}}\prob(D = 0\,|\, X, Y)=\frac{n_0}{\# \{D = 0\}} -\frac{n_0}{\# \{D = 0\}}e^{-v^2/2},
\end{align*}
where $ v = X^\top \gamma_x + Y \gamma_y $, $n_1$ ($n_0$) are the number of cases (controls) in the sample respectively, and $\#\{D = 1\}$ ($\#\{D = 0\}$) is the number of cases (controls) in the population respectively. 

Adding these two terms together, we obtain
\begin{align*}
    Q_y(x)\propto \frac{n_0}{\{\# D = 0\}} \N(x;0,\Sigma) + \left\{\frac{n_1}{\{\# D = 1\}}  - \frac{n_0}{\{\# D = 0\}}\right\} \N(x;0,\Sigma)  e^{-v^2/2}. 
\end{align*}
Here, $\prob(x) = \N(x;0,\Sigma) $ denotes the density of a multivariate Gaussian with mean $0$ and covariance $\Sigma$ evaluated at $x$. After expanding the squares, the second term can be written as 
\begin{align*}
   \N(x;0,\Sigma)  e^{-v^2/2} &=  \frac{1}{(2\pi)^{p/2}|\Sigma|^{1/2}} \exp\left\{-\frac{1}{2} \left(x^\top\Sigma^{-1}x + x^\top \gamma_x\gamma_x^\top x + 2\gamma_y y x^\top \gamma_x + \gamma_y^2 y^2\right)\right\}\\
   & =  \frac{1}{(2\pi)^{p/2}|\Sigma|^{1/2}} \exp\left\{-\frac{1}{2}(x-\tilde{\mu})^\top\tilde{\Sigma}^{-1}(x-\tilde{\mu})+\frac{1}{2} \gamma_y^2y^2(\gamma^\top\tilde{\Sigma}\gamma - 1)\right\},
\end{align*}
where $\tilde{\Sigma} = (\Sigma^{-1} + \gamma_x \gamma_x^\top)^{-1}$, $\tilde{\mu} = -\tilde{\Sigma}\gamma_y y \gamma_x $. Thus,
\begin{align*}
     \N(x;0,\Sigma)  e^{-v^2/2} = \frac{|\tilde{\Sigma}|^{1/2}}{|\Sigma|^{1/2}}\exp\left(\frac{1}{2} \gamma_y^2y^2(\gamma_x^\top\tilde{\Sigma}\gamma_x - 1)\right)\N(x;\tilde{\mu},\tilde{\Sigma}).
\end{align*}
From here, we see that $Q_y(x)$ is approximately a mixture of two Gaussian distributions
\begin{equation*}
    f_1(x) = \N(x; \tilde{\mu},\tilde{\Sigma}),\quad f_2(x) = \N(x; 0,\Sigma),
\end{equation*}
and the (relative) weights are 
\begin{equation*}
    w_1 = \left\{\frac{n_1}{\#\{ D = 1\}}  - \frac{n_0}{\# \{D = 0\}}\right\}\frac{|\tilde{\Sigma}|^{1/2}}{|\Sigma|^{1/2}}\exp\left(\frac{1}{2} \gamma_y^2y^2(\gamma^\top\tilde{\Sigma}\gamma - 1)\right),\quad w_2 = \frac{n_0}{\# \{D = 0\}}. 
\end{equation*}
To generate tilted knockoff for an observation $x$ from this Gaussian mixture, we follow steps in \cite{gimenez2019multipleKO}. First, we compute the probability of $x$ being in the first mixture given by
\begin{equation}
    q_1 = \frac{w_1 f_1(x)}{w_1 f_1(x) + w_2 f_2(x)} =  \frac{w_1'e^{-\frac{1}{2}(x-\tilde{\mu})^\top\tilde{\Sigma}^{-1}(x-\tilde{\mu})}}{w_1'e^{-\frac{1}{2}(x-\tilde{\mu})^\top\tilde{\Sigma}^{-1}(x-\tilde{\mu})} + w_2 e^{-\frac{1}{2}x^\top\Sigma^{-1}x}},
\end{equation}
where $w_1' = \left\{\frac{n_1}{\{\# D = 1\}}  - \frac{n_0}{\{\# D = 0\}}\right\}\exp\left(\frac{1}{2} \gamma_y^2y^2(\gamma_x^\top\tilde{\Sigma}\gamma_x - 1)\right)$. Next, we sample a Bernoulli variable $Z\sim \mathrm{Bernoulli}(q_1)$. If $Z = 1$, we sample the knockoffs pairwise exchangeable w.r.t. $f_1$ and otherwise $f_2$.

%Because we \textit{fix} $y$ at the observed value, $Q_y(x)$ is the \textit{conditional} distribution of $x\,|\,y$ of $\N(0,\Xi)$. Comparing with Eqn.\eqref{eq:Theta}, Eqn.\eqref{eq:Xi} does not involve $\beta$, and therefore we do not need to know $\beta$ to be able to construct kknockoffs.

\section{Other methods to sample approximate tilted knockoffs}

\subsection{Deep knockoff} \label{appendix:deep}

The deep knockoff method, proposed in \cite{romano2020deep}, does not require an analytic formula to generate knockoffs. Instead, it directly produces a knockoff generating mechanism that satisfies the exchangeability criteria \eqref{eq:exchangeability_q}. The deep knockoff method takes three inputs. First, a training samples $X$; for our application, the training samples are i.i.d. sampled from $Q_y(X)$. Second, a deterministic function $f_{\theta}(X, V)$ from which to sample knockoffs. Here, $\theta$ is a parameter which we optimize over, and $V$ is a new variable. \cite{romano2020deep} defines $f_{\theta}$ to be a neural network to allow for flexible knockoff distributions. Third, a score function to evaluate whether the generated knockoff variables approximately satisfy the exchangeability condition w.r.t. $Q_y(X)$. The score function includes three components: the maximum mean discrepancy between the distributions $(X, \tilde{X})$ and $(X, \tilde{X})_{\text{swap}(X_j,\tilde{X}_j)}$, discrepancy between the two moments of $X$ and $\tilde{X}$, and the correlations between $X_j$ and $\tilde{X}_j$. The deep knockoff method optimizes the parameter $\theta$ to minimize the score function. 

Because the tilted distribution $Q_y(X)$ depends on the response variable $Y$, we need to train one deep knockoff machine for each distinct $Y$ values, and this can be computationally intensive. Thus, to illustrate this method, we consider an example where $Y$ only takes two values $Y\in \{0, 1\}$, and $Y\,|\, X$ is from a logistic model. This example has the same set up as Appendix~\ref{appendix:so_estimated}, with the following exceptions: the number of variables is $p=50$. The number of non-null $\beta$ and $\gamma_x$ are both 20 and the non-null variables are i.i.d. $\N(0,0.2^2)$. We use rejection sampling to generate training samples: if $Q_y(x) \leq \prob(x) M$ for some $M<\infty$, then we can sample $X\sim \prob$ and then accept with probability $Q_y(x)/\prob(x) M$. Since 
\begin{equation*}
    Q_y(x) = \frac{1}{M}\prob(x) \prob(S=1\,|\,x, y)\leq \frac{1}{M}\prob(x) 
\end{equation*}
where $M$ is the normalizing constant and the last inequality is because $\prob(S=1\,|\,x, y) < 1$, we  accept with probability $\prob(S = 1\,|\, X, y)$. 

\begin{figure}[ht]
    \centering
    \includegraphics[width = 1\textwidth]{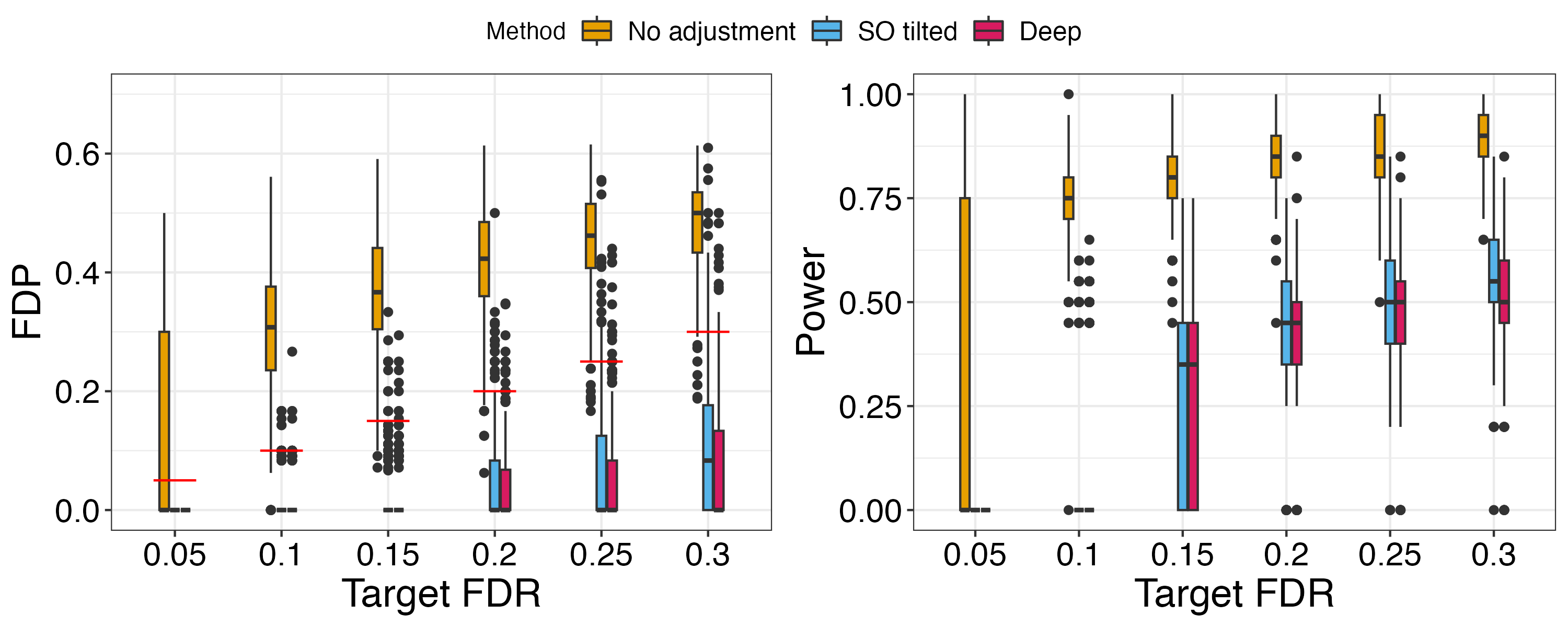}
    \caption{Boxplot of the FDP and power in $B = 500$ repetitions at each target FDR level using the deep knockoff machine for example in Appendix~\ref{appendix:deep}. The target FDR level is shown in red segments.}
    \label{fig:deep2}
\end{figure}

We plot the FDP using the deep knockoff machine in Figure~\ref{fig:deep2}. The deep knockoff machine controls the FDR and it has comparable performance as the second order tilted knockoffs (Figure~\ref{fig:deep2}).  

\subsection{Metropolized knockoff}\label{appendix:metro}

The metropolized knockoff algorithm allows researchers to sample knockoffs for observations from any known sample distribution \cite{bates2021metro}. Metropolized knockoff is motivated by the observation that suppose two variables $(X, \tilde{X})$ are two consecutive terms on a reversible Markov chain (MC), then their joint distribution is exchangeable. In particular, \cite{bates2021metro} use the Metropolis-Hastings algorithm to construct such a reversible MC. For example, to sample knockoffs for the first variable $X_1$, we sample $\tilde{X}_1$ from a reversible MC whose stationary distribution is $\law(X_1\,|\, X_{-1})$. We start by sampling a proposal $X_1^\star$ from $q(\cdot\,|\, X_1)$ and then we accept the proposal (i.e., set $\tilde{X}_1 = X_1^\star$ and otherwise set $\tilde{X}_1 = X_1$) with probability 
\begin{equation*}
    \prob(\text{accept} X_1^\star\,|\, X) = \min\left(1, \frac{\law(X_1^\star \,|\,X_{-1}) q(X_1\,|\, X_1^\star)}{\law(X_1 \,|\,X_{-1}) q(X^\star_1\,|\, X_1)}\right).
\end{equation*}
In order to generate knockoffs for all of the variables, this process is applied sequentially, at each variable $j$ setting the stationary distribution to be $\law(X_j\,|\, X_{-j}, \tilde{X}_{1:(j-1)},X_{1:j}^\star)$. Algorithm \cite[Algorithm~1]{bates2021metro} summarizes the metropolized knockoff algorithm.

Computing the acceptance probability is typically computationally intensive, as it requires $O(2^p)$ computations in the most general scenario when the number of variables is $p$. \cite{bates2021metro} developed an efficient sampling procedure when the joint distribution of $X$ can be represented as a graphical model. In particular, if we construct a junction tree $T$ from the graph, then the computation complexity is $O(p2^w)$ where $w$ is the tree width of $T$. 

\begin{figure}[ht]
    \centering
    \includegraphics[width = 1\textwidth]{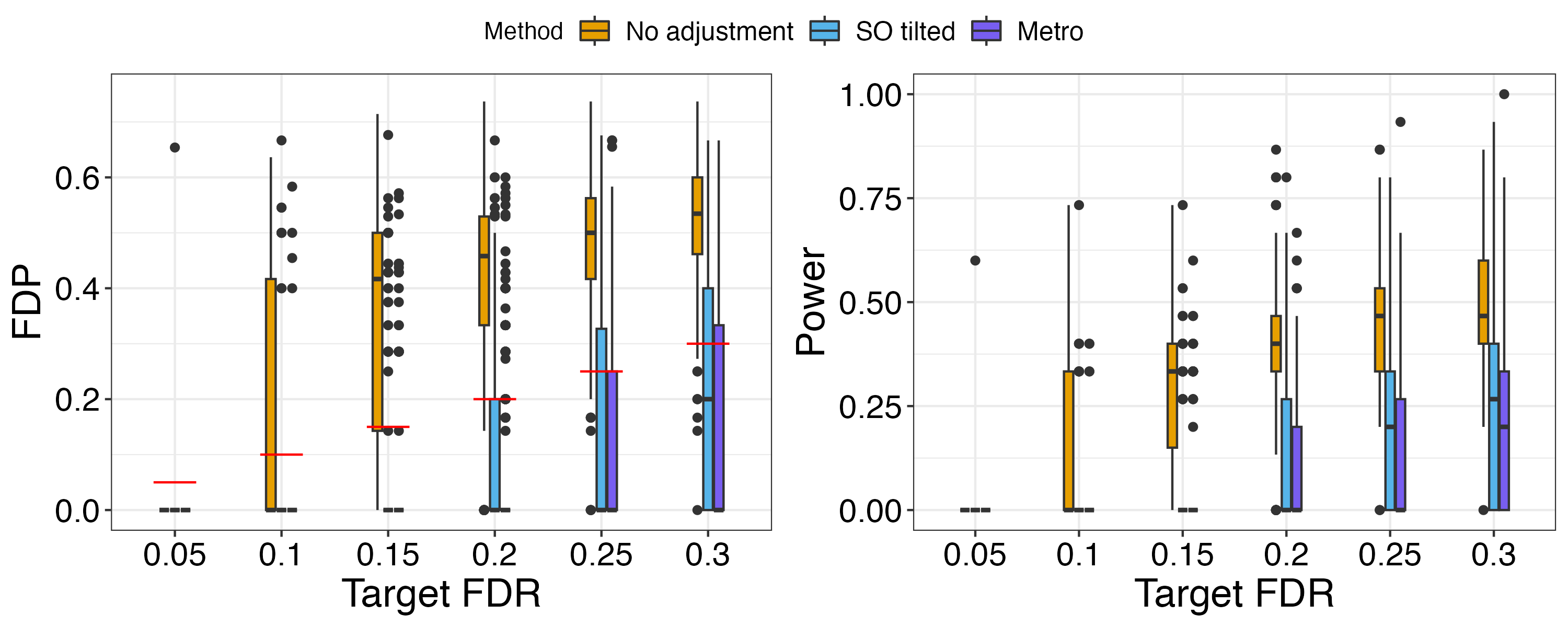}
    \caption{Boxplot of the FDP in $B = 250$ repetitions at each target FDR level using the metropolized knockoff method for example in Appendix~\ref{appendix:metro}. The target FDR level is shown in red segments.}
    \label{fig:metro}
\end{figure}

We apply the metropolized knockoff procedure to the same setting as in Section~\ref{section:second_order_tilted} with two modifications. First, only 15 variables are non-nulls for both $\beta_j$ and $\gamma_j$ and the non-null variables are sampled from $\N(0, 0.25^2)$. The smaller number of non-nulls reduces the computation to sample knockoffs and memory cost to store acceptance probabilities. Second, the non-null variables among $\beta_j$ and $\gamma_j$ do not overlap. Now the FDR of standard model-X knockoff is strongly inflated (Figure~\ref{fig:metro}) compared with previous settings where the non-nulls are allowed to overlap. Third, we set $\gamma_y = -4$. Figure~\ref{fig:metro} shows that sampling tilted knockoff using either second order approximation or metropolized knockoff controls the FDR. In particular, metropolized knockoff is quite conservative (the average FDP is 0.15 at target FDR level $q = 0.3$). Both second order and metropozlied tilted knockoffs are less powerful compared to the standard knockoff, and the two methods to sample tilted knockoffs yields comparable power.

\section{Additional results from Section~\ref{section:real}}

In this section, we report additional details of knockoff generation and results of the estimated probability of diagnosis $D$ given genotype $X$ and secondary phenotype $Y$. 

\subsection{Estimated probability of diagnosis}

We report the pseudo-$R^2$ defined as 
$$
\text{pseudo-}R^2 = \frac{\text{null deviance} - \text{deviance}}{\text{null deviance}}
$$
in Table~\ref{tab:pseudoR2}. 

\begin{figure}[t]
\centering
   \includegraphics[width = 0.8\textwidth]{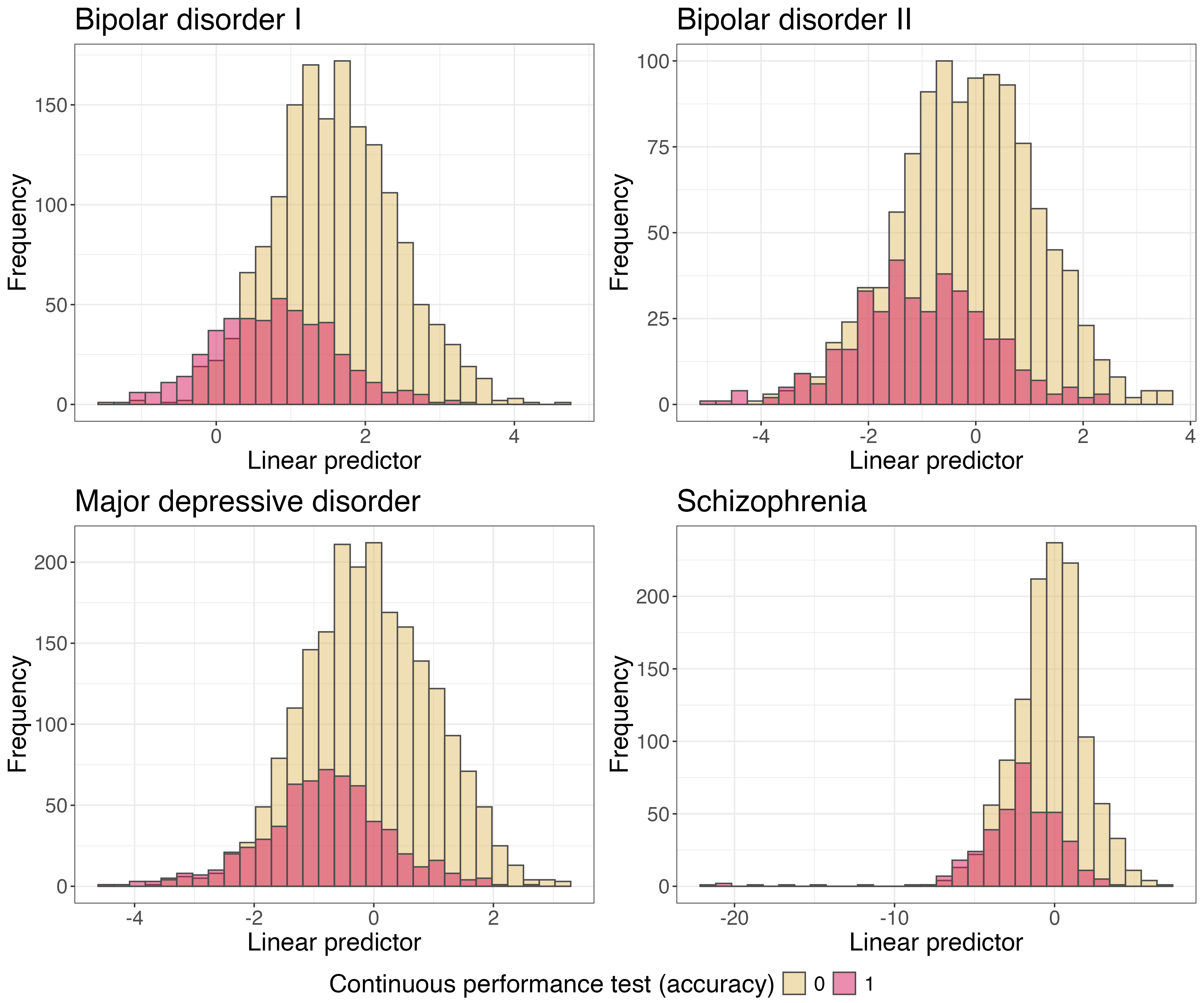}
   \caption{Linear predictor in the logistic regression for each diagnosis that comes from genetic variants when the test score of continuous performance test (accuracy) is among the top 20\% quantiles (red) or the remaining 80\% quantiles (yellow). }
    \label{fig:att_a_linear_predictor}
\end{figure}

Figure~\ref{fig:att_a_linear_predictor} shows the contribution of genetic variants in the linear predictor of the logistic regressions (i.e., $X^\top \gamma_x$) when the secondary phenotype $Y$ is accuracy in the continuous performance test. Figures for the other secondary phenotypes are similar. In general, higher neurocognitive function leads to a lower probability of each diagnosis. 

\begin{table}[ht]
    \centering
    \begin{tabular}{|c | c |c c c c | }
    \hline
   &  & \multicolumn{4}{c|}{Diagnosis} \\
\cline{3-6}
    \multirow{3}{*}{Phenotype} & \multirow{3}{*}{Measure}    &      Bipolar    & Bipolar  & \multirow{2}{*}{Schizophrenia} & Major Depressive \\
&     &  Disorder I & Disorder II & &  Disorder \\
     \hline
  \multirow{2}{*}{Attention}&   Accuracy    & 0.11 & 0.20& 0.35 & 0.16 \\
    & Speed & 0.07 & 0.18 & 0.30  & 0.14\\
    \hline
       Processing & Accuracy & 0.13& 0.01& 0.36 & 0.14\\
      speed & Speed & 0.06 & 0.02 & 0.38 & 0.14\\
      \hline
    \end{tabular}
    \caption{Psuedo-$R^2$ of the fitted logistic regression using individual secondary phenotypes (rows) and genotypes to estimate probability of each diagnosis (columns).  }
    \label{tab:pseudoR2}
\end{table}

\subsection{Details of knockoff generation}\label{app:real_knockoff_generation}
We use multiple knockoffs of size 3 as proposed in \cite{gimenez2019multipleKO} to obtain more stable selections. Correlation between each variable and its knockoffs are obtained using both the maximum entropy (ME) \cite{gimenez2019multipleKO} and minimizing reconstructability (MVR) knockoffs \cite{spector2022KOpower} as they have been shown be more powerful compared to the equi-correlation and SDP knockoffs. We generate 50 knockoffs of each type, and thus we obtain 100 sets of selected variables in total for the secondary phenotype, where 50 derived from from ME and 50 from MVR knockoffs. We use the maximum penalty parameter at which a variable enters the fitted model in a regularized logistic regression as the feature importance statistics.  

\printbibliography
\end{document}